\documentclass[times]{stvrauth}

\usepackage{moreverb}

\usepackage{color,soul}

\usepackage[colorlinks,bookmarksopen,bookmarksnumbered,citecolor=red,urlcolor=red,driverfallback=dvipdfm]{hyperref} 

\newcommand\BibTeX{{\rmfamily B\kern-.05em \textsc{i\kern-.025em b}\kern-.08em
T\kern-.1667em\lower.7ex\hbox{E}\kern-.125emX}}

\def\volumeyear{2010}

\usepackage{graphicx}
\usepackage{listings}
\usepackage{adjustbox}
\usepackage{fancybox}
\usepackage{subcaption}

\usepackage[table]{xcolor}

\makeatletter
\newenvironment{CenteredBox}{%
\begin{Sbox}}{
\end{Sbox}\centerline{\parbox{\wd\@Sbox}{\TheSbox}}}
\makeatother

\lstdefinestyle{Scala}
{ 
	language=scala,
	basicstyle = \footnotesize\sffamily,
	showspaces = false,
	showstringspaces = false,
	showtabs = false,
	tabsize = 1,
	breaklines = true,
	breakatwhitespace= false,
	numbers = left,
	columns=fullflexible,
	sensitive=true,
	string=[b]",
	morestring=[d]',
	morecomment=[l]{//},
}

\usepackage{xcolor}

\usepackage{array}
\usepackage{multirow}

\usepackage{setspace}

\newcommand*{\transmut}{\textsc{TRANSMUT-Spark}}

\newcommand*{\courier}{\fontfamily{pcr}\selectfont}
\newcommand{\chl}[1]{{\courier \textbf{#1}}}
\newcommand{\ct}[1]{{\courier #1}}

\begin{document}

\runningheads{J.~B. de Souza Neto et al.}{\textsc{TRANSMUT-Spark}: Transformation Mutation for Apache Spark}

\title{\textsc{TRANSMUT-Spark}: Transformation Mutation for Apache Spark}

\author{Jo\~{a}o Batista de Souza Neto\affil{1}\corrauth,
Anamaria Martins Moreira\affil{2}, Genoveva Vargas-Solar\affil{3} and Martin A. Musicante\affil{1}}

\address{\affilnum{1}Department of Informatics and Applied Mathematics (DIMAp), Federal University of Rio Grande do Norte (UFRN), Natal, Brazil.\break
\affilnum{2}Computer Science Department (DCC), Federal University of Rio de Janeiro (UFRJ), Rio de Janeiro, Brazil.\break
\affilnum{3} French Council of Scientific Research (CNRS), LIRIS, Lyon, France.}

\corraddr{Jo\~{a}o Batista de Souza Neto, Department of Informatics and Applied Mathematics (DIMAp), Federal University of Rio Grande do Norte (UFRN), Natal, Brazil. E-mail: jbsneto@ppgsc.ufrn.br}

\begin{abstract}

We propose \textsc{TRANSMUT-Spark}, a tool that automates the mutation testing process of Big Data processing code within Spark programs.
Apache Spark is an engine for Big Data Analytics/Processing.
It hides the complexity inherent to Big Data parallel and distributed programming and processing through built-in functions, underlying parallel processes, and data management strategies. 
Nonetheless, programmers must cleverly combine these functions within programs and guide the engine to use the right data management strategies to exploit the large number of computational resources required by Big Data processing and avoid substantial production losses. 
Many programming details in data processing code within Spark programs are prone to false statements that need to be correctly and automatically tested.
This paper explores the application of mutation testing in Spark programs, a fault-based testing technique that relies on fault simulation to evaluate and design test sets.
The paper introduces the \textsc{TRANSMUT-Spark}   solution for testing Spark programs. 
\textsc{TRANSMUT-Spark} automates the most laborious steps of the process and fully executes the mutation testing process. 
The paper describes how the tool automates the mutants generation, test execution, and adequacy analysis phases of mutation testing with \textsc{TRANSMUT-Spark}. 
It also discusses the results of experiments that were carried out to validate the tool to argue its scope and limitations.

\end{abstract}


\maketitle


\section{Introduction}

In recent years (i.e., the Big Data phenomenon), data production has introduced a series of challenges for its storage and processing using greedy algorithms. 
These algorithms call for well-adapted processing infrastructures and programming models to support their execution on clustered like architectures to be implemented on large-scale conditions. 
Existing frameworks adopt either control flow~\cite{hadoop273,katsifodimos2015apache} or data flow approaches~\cite{yu2008,beam2016,Zaharia:2010}. 
Apache Spark~\cite{Zaharia:2010} has emerged as one of the main engines for Big Data processing because it screens difficulties inherent to parallel programming and distributed processing to programmers, allowing them to focus only on the algorithmic aspects of Big Data processing programs. 
Spark adopts a data flow program model with programs represented as directed acyclic graphs (DAGs) that coordinate the execution of operations applied on datasets distributed across several computing nodes.

Even if Spark promotes a less painful way of designing, programming, and executing big data parallel processing programs, programmers still need to tune several aspects within their code and resource allocation configuration. 
This tuning is not an easy task to do in the design and programming phases due to many aspects. 
Therefore, testing this type of program becomes essential to avoid losses in production~\cite{Garg2016}.
In this context, software testing techniques emerge as relevant tools~\cite{Meeker2014,LIU2016134}. 
This paper addresses the testing of big data processing code weaved within programs by exploring the application of \textit{mutation testing}~\cite{DeMillo1978} on Apache Spark programs.

Mutation testing is a powerful test development technique where tests are designed to pinpoint specific faults introduced in the code, the \emph{mutations}. The quality of the resulting tests is closely dependent on how representative these faults are concerning the programs developed in that specific language or programming paradigm. Some kinds of faults, such as using the wrong Boolean operation in a logic expression, are common to most languages. Others, however, are specific to the language or paradigm.
In a previous paper~\cite{caise2020} we proposed a  \textit{transformation mutation} approach that applies mutation testing in Spark programs introducing mutation operators designed to simulate faults specific to data flow programs. 
We proposed 15 mutation operators divided into two groups: \textit{data flow} and \textit{transformations}. 
Data flow operators model faults by modifying the program's data flow graph, while operators for transformations modify specific data operations (called transformations). 
In our previous work~\cite{caise2020}, we conducted manual experiments to show our mutation operators applicability to test Spark programs.
Even if results were promising, we confirmed that mutation testing could be laborious, time-consuming and prone to errors~\cite{Maldonado2001} if it is not partially automated. We believe that mutation testing must be delegated to a tool to perform intensive testing and face software development conditions that imply production speed and quality requirements. 
Therefore, we propose \transmut, a tool that automates the Spark programs' mutation testing process.

\transmut\ implements the most common requirements that a mutation test tools must provide~\cite{delamaro1996proteum}: \textit{test case handling}, which includes the execution, inclusion, and exclusion of test cases; \textit{mutant handling}, for generating, selecting, executing and analyzing mutants; and \textit{adequacy analysis}, which involves calculating the mutation score and generating reports. 
The paper shows how \transmut\ automates the primary and most laborious steps of the process and makes it possible to fully execute the mutation testing process in Spark programs.
These features are thoroughly validated through experimental settings discussed in the paper. 
\transmut\ was used to perform a new, complete round of experiments of the mutation approach proposed previously~\cite{caise2020}.  
Additionally, \transmut\ was also experimentally compared with traditional mutation testing done at the (syntactic) programming language level. Experiments showed that \transmut\ is complementary to mutation testing for Scala programs that do not address  Big Data processing code. Thus, combining mutation testing approaches addressing programming languages statements with \transmut\  can test both higher-level Big Data processing code that is weaved within classic imperative programs and the more basic program constructs.
The contribution presented in this paper is twofold: (1) \transmut for automating mutation testing of Big Data processing code weaved within Scala programs, (2) and an experimental battery of tests evaluating the possibilities of mutation testing for Spark programs that could not be carried out without the tool.

Accordingly, the remainder of the paper is organized as follows. 
Section \ref{sec:background} introduces the background concepts behind \transmut{} namely Apache Spark and mutation testing. Section \ref{sec:related} introduces related work concerning approaches addressing the problem of testing programs implemented to run on top of big data processing platforms. Section \ref{sec:mutation_operators} introduces the set of mutation operators we propose for Apache Spark programs. Section \ref{sec:transmut} introduces \transmut{} the tool that we propose for automating the process of testing Spark programs written in Scala. Section \ref{sec:experiments} describes the experiments that we conducted for evaluating \transmut{}  as a testing tool. 
Section~\ref{sec:transmut-vs-scalamu} describes a comparative study on transformation mutation and traditional mutation testing.
Section~\ref{sec:threats} discusses the limitations and threats to validity.
Finally Section \ref{sec:conclusion} concludes the paper and discusses future work.

\section{Background}\label{sec:background}
The approach behind \transmut{} relies on two conceptual underpinnings: Apache Spark and mutation testing. This section introduces the fundamental concepts of these components underlying those that are key to understand the transformation mutation strategy for Big Data processing code in Spark promoted by \transmut{}.

\subsection{Apache Spark}\label{subsec:spark}

\textit{Apache Spark}~\cite{Zaharia:2010} is an execution platform for large-scale data processing parallel programs written in programming languages like Scala, Java, Python, and R. 
It is suited for embedding machine learning algorithms and interactive analysis, such as exploratory queries on datasets within programs.
As such, it offers  libraries for working with structured data using an SQL-style API (\textit{SparkSQL}), machine learning (\textit{MLlib}), streaming data processing (\textit{Spark Streaming}) and graph processing (\textit{GraphX})~\cite{spark22}.

The objective of Spark is to make parallel programming transparent. It prevents programmers from dealing with data distribution among cluster nodes and ensuring fault tolerance and processes' synchronization, similar to classic parallel programming approaches. For dealing with fault-tolerant data distribution, Spark is based on a data abstraction called {\em Resilient Distributed Dataset} (RDD)~\cite{Zaharia:2012}. An RDD represents a collection of data distributed through the nodes of a cluster that can be processed in parallel within a Spark program. 
In the case of distributed data collections as an RDD, fault-tolerant means that its partitions can be reconstructed on the emergence of failures in the execution process. 
Therefore, Spark uses a timeline system that tracks the sequence of operations that must be performed to reconstruct a partition when data is lost.

\begin{figure*}
   \centering
   \includegraphics[width=\linewidth]{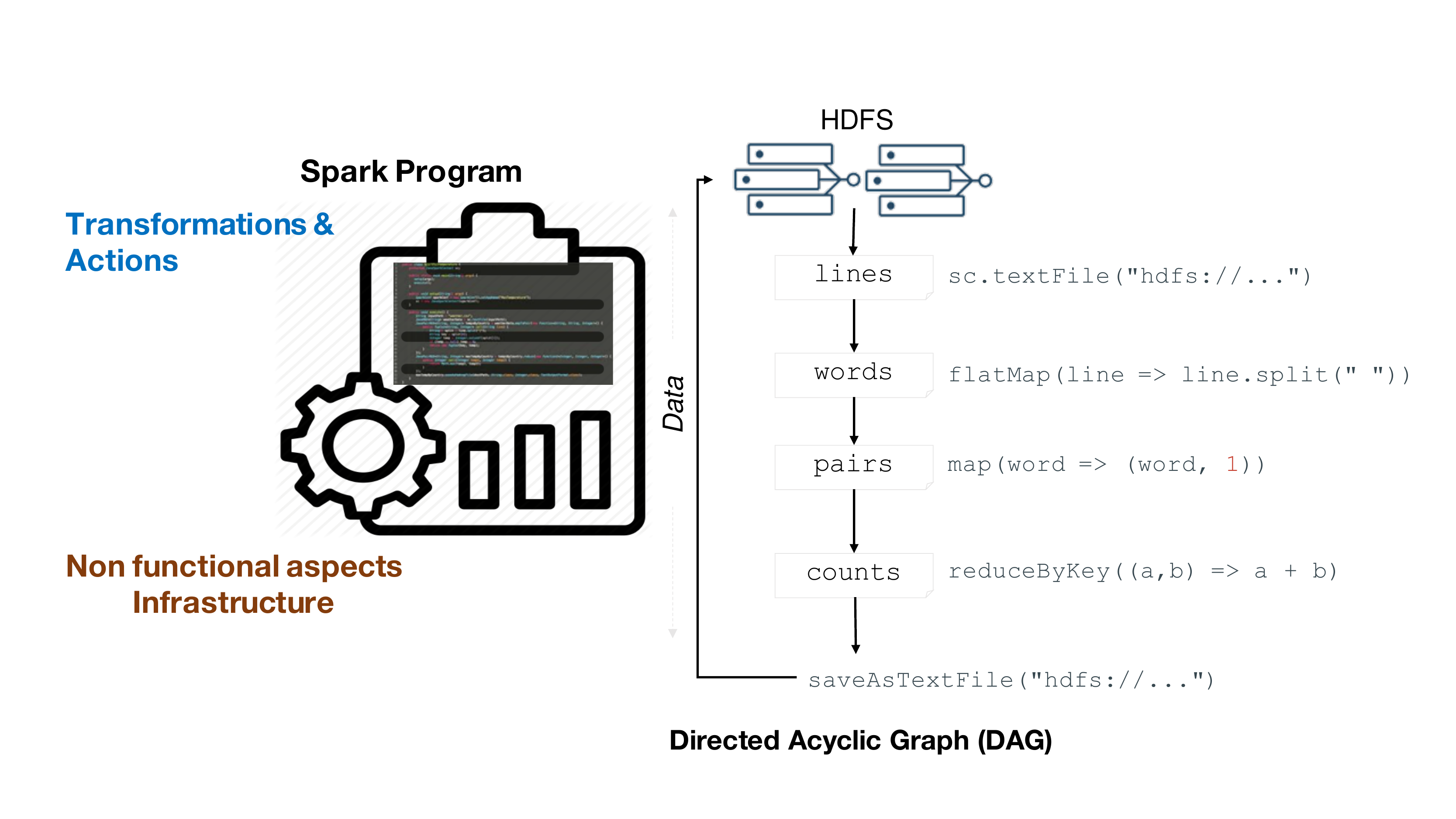}
 \caption{Spark program overview.}
   \label{fig:spark-program}
 \end{figure*}

The general principle of a Spark program is shown in Figure \ref{fig:spark-program}. It consists of two types of blocks: 
(1) functional aspects expressed by a data flow implementing the application logic;  
(2) operations devoted to process data (transformations) and to guide the way data is managed from memory to cache and disk (actions).

As shown in Figure \ref{fig:spark-program}, the data used by a program is initially stored in persistence support, often in a distributed file system, for example, on HDFS (Hadoop Distributed File System)~\cite{hadoop273}. The data are first retrieved from the persistence support and transformed into an RDD for executing a Spark program. The data processing operations stated in a Spark program will then be applied to this RDD.
For example, the program in the figure is intended to count the occurrence of words in a text dataset. The program runs through the dataset to first separate the lines of text into separate words. So, for each word, a key/value tuple is created containing the word as key and the integer $1$ as value. Afterwards, the key/value tuples are grouped by the key, and the values are aggregated, resulting in a dataset with the words and their frequencies, \textit{i.e.}, the number of times it appears in the text. Finally, the resulting dataset is stored in a storage system (HDFS). The sequence of operations applied in a Spark program forms a DAG (Directed Acyclic Graph). This DAG defines the data flow of a Spark program and represents the program's execution plan that will be optimized when executed in a cluster.

In the program there are two  classes of Spark operations: {\em transformations} and {\em actions}. 
Understanding the way the Spark platform executes them 
is essential to understand the implications of using them within a program.

\paragraph*{Transformations:} 
create new RDDs from existing ones~\cite{spark22}. 
In Spark, transformations are evaluated under a \textit{lazy evaluation}~\cite{Zaharia:2010} strategy \footnote{In programming language theory, lazy evaluation, or call-by-need, is an evaluation strategy that delays the evaluation of an expression until its value is needed (non-strict evaluation) and which also avoids repeated evaluations (sharing).}, so that transformations are not executed when are called in a program. However, their execution is deferred until other computations need them.
 Spark transformations can receive functions as input parameters and apply these functions over the elements of RDDs to generate the elements of a new RDD.

Spark's transformations can be divided according to the type of processing that is done: mapping transformations (such as \textit{map} and \textit{flatMap}), which apply a function to map the values of an input RDD to an output RDD; filtering transformations, which filter the values of an RDD based on a predicate function; grouping  transformations (\textit{groupBy}, \textit{groupByKey}), which group the values of an RDD based on a key; aggregation transformations (\textit{reduceByKey}, \textit{aggregateByKey}), which aggregate values grouped by a key; set transformations (\textit{union}, \textit{intersection}, \textit{subtract}, \textit{distinct}), which are similar to mathematical set operations; join transformations (\textit{join}, \textit{leftOuterJoin}, \textit{rightOuterJoin}, \textit{fullOuterJoin}), which join two RDDs on the basis of a common key; and sort transformations (\textit{sortBy}, \textit{sortByKey}), which sorts the values in an RDD.

Transformations can be either \textit{narrow} or \textit{wide}, depending on the degree of distribution of the data sets~\cite{Zaharia:2012}. 
Narrow transformations are those where each partition of the original RDD is only used by a maximum of one partition of the new RDD generated by the operation. 
Operations like \textit{map}, \textit{flatMap} and \textit{filter} are examples of narrow transformations. 

Wide transformations are those where each partition of the original RDD is used to create multiple partitions of the new RDD. 
Generally, this type of transformation operates on key/value tuple datasets that need to have their data grouped by key, such as \textit{reduceByKey} and \textit{join} operations. 
This type of operation causes a reorganization of the data in the cluster to group values that have the same key to the same partition. 
This process of data reorganization is called \textit{shuffling}. 
This mechanism involves copying and sending data between the cluster nodes, making it a complex and costly process.

\paragraph*{Actions:} 
return values that are \textit{not} RDDs to the Driver or write the content of the RDD in some storage system. 
Since actions generate concrete results, these are the operations that trigger the lazy evaluation execution of transformations. In this way, Spark can optimize the execution of the applications by executing narrow transformations in the same process (pipeline) and create different stages for wide operations that trigger the shuffling process~\cite{Zaharia:2012}.

Some of the main Spark actions are \textit{reduce}, which aggregates all the values of an RDD into a single value; \textit{collect}, which returns to the Driver the contents of an RDD in the form of an array; and \textit{saveAsTextFile}, which saves the content of an RDD in a storage system.

A Spark program execution is coordinated by a \textit{Driver}  through  a  \textit{SparkContext}. The \textit{SparkContext}  executes the main program function consisting of a sequence of operations on RDDs and manages internal program information and deploys the operations on  the cluster nodes~\cite{ganelin2016spark}.

\paragraph*{Spark program example} 
Figure~\ref{fig:wordcount-example} shows the code of the counting words example programmed in Apache Spark using  Scala as the programming language. 
The program starts by reading a data set and storing it in the \textit{input} variable (line 1).
Next, it applies the \textit{flatMap} transformation, which separates each line of text into words (line 2). 
The RDD with words is transformed into a key/value RDD with the application of the transformation \textit{map} which generates key/value pairs in which the key is a word, and the value is the integer number 1 (line 3). 
The counting of words is done with the application of the \textit{reduceByKey} transformation, which groups the values per key and then applies the (associative) addition function to sums all the values, resulting in the RDD \textit{count}, containing the frequency of each word (line 4). 
The program is terminated by calling the action \textit{saveTextFile} which performs the transformations and saves its result (line 5).

\begin{figure}
\centering
\begin{CenteredBox}
\begin{lstlisting}[style=Scala]
val input = sc.textFile("hdfs://...")
val words = input.flatMap( (line: String) => line.split(" ") )
val pairs = words.map( (word: String) => (word, 1) )
val counts = pairs.reduceByKey( (a: Int, b: Int) => a + b )
counts.saveAsTextFile("hdfs://...")
\end{lstlisting}     
\end{CenteredBox}
\caption{Example of a Spark program.}
\label{fig:wordcount-example}
\end{figure}

A particular characteristic of a Spark program is that it includes operations devoted to processing data (application logic) and operations devoted to guiding the way data are managed to execute the program 
(\textit{i.e.}, data swapped from disk to cache and memory, shared or replicated across different processes and nodes, transmitted across processes). Data management decisions are independent of the application logic but have a significant impact on its performance.
Designing a Spark program implies ensuring functional faults are inconsistencies within the code that may result in unexpected behaviour when executed, for example, wrong transformations, incorrect parameters, and absence of transformations. In this paper we focus on these types of faults (see Section \ref{sec:mutation_operators}).

\subsection{Mutation Testing}\label{subsec:mutation_testing}
De~Millo \textit{et al.}~\cite{DeMillo1978} proposed {\em mutation testing},  a fault-based testing technique that consists of creating variants of a program and then deriving tests that can show that the variants behave differently from the original program.
These variants, called {\em mutants}, are obtained from the program to be tested by performing minor modifications that simulate common faults or programmers mistakes.
These modifications are systematically created by using predefined rules, called \textit{mutation operators}. 
Mutation operators insert faults in the program and generate one or more mutants of a given source program.

Mutation testing establishes mutants as test requirements so that tests must be designed to distinguish mutants from an original program. 
In other words, a test should identify that the result obtained with a mutant is different from the result obtained with the original program. 
When this occurs, the test is said to \textit{kill} the mutant. 
It is impossible to kill a mutant in certain situations because it has the same behaviour as the original program and cannot be distinguished. In this case, the mutant is said to be \textit{equivalent}, and it is removed from the test requirements set.

The test developer can then use each non-equivalent mutant as a guide to derive interesting test cases
(\textit{i.e.}, input test data that will exercise the program in such a way that if each of those specific faults were to be in the code, there would be a test case in the test set that is capable of showing its presence to the tester). A typical example is the case o a mutant where a conditional construct guarded by a comparison of the kind $a < b$ is changed into $a\leq b$. Unless we have a test case where $a$ and $b$ have the same value at this point, the mistake will remain unnoticed.

Mutation testing coverage as a testing requirement can be determined from the ratio of the number of killed mutants to the total number of mutants, not considering equivalent mutants. 
This ratio is known as {\em mutation score}, being used as a quality measure for test sets~\cite{offut:2010}. 
The following formula calculates the mutation score:
\begin{align*}
\begin{split}
ms(P, T) ={}& \dfrac{DM(P, T)}{M(P) - EM(P)}
\end{split}
\end{align*}

\noindent where $ ms(P, T) $ is the mutation score for a program $ P $ and a test set $ T $, $ DM(P, T) $ is the number of mutants of the program $ P $ killed by the test set $ T $, $ M(P) $ is the number of mutants generated from $ P $, and $ EM(P) $ is the number of mutants equivalent to $ P $.

\begin{figure}[!htbp]
	\centering
	\includegraphics[width=\textwidth]{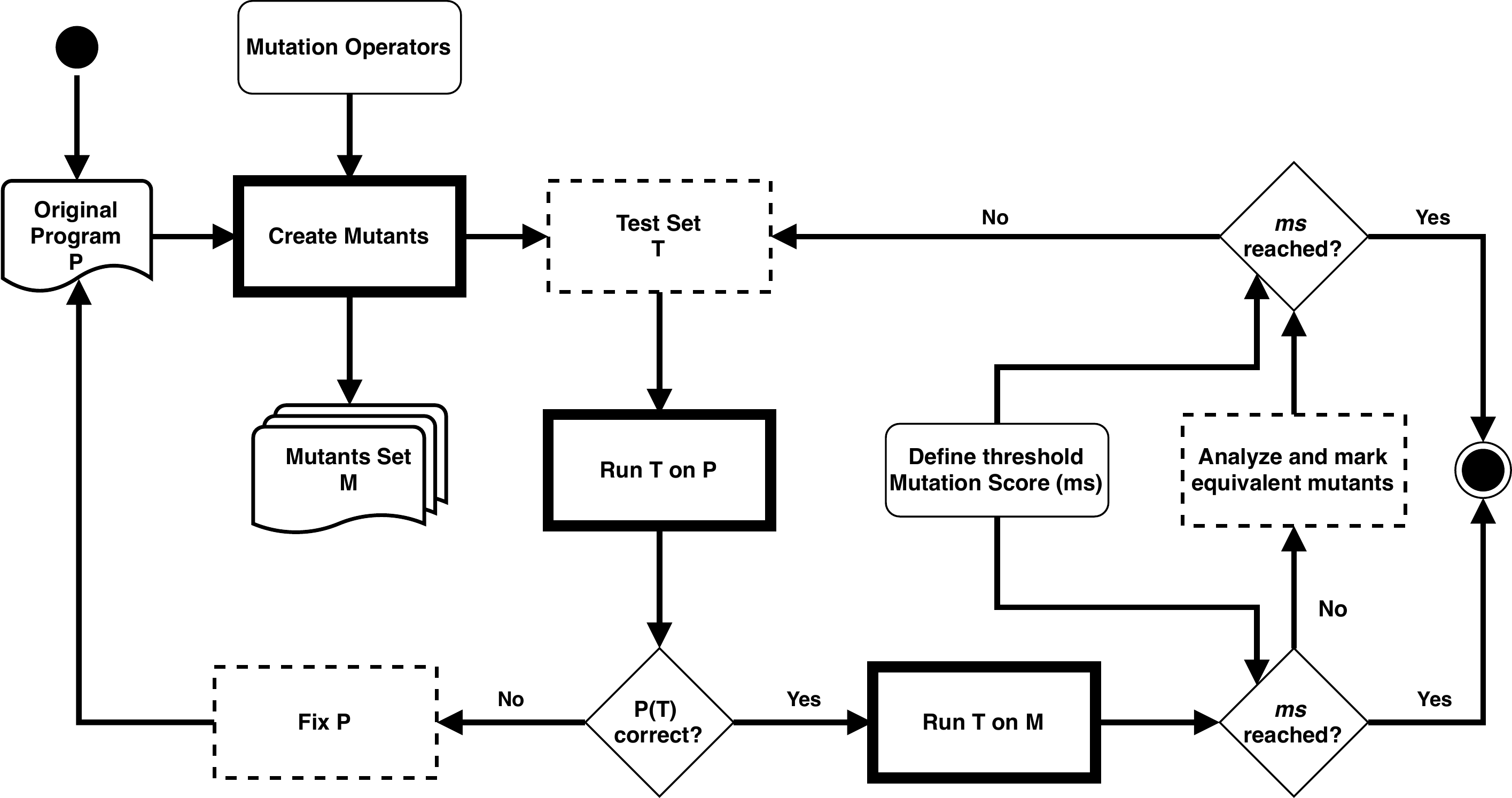}
	\caption{Mutation testing process (adapted from \cite{offut:2010}).}
	\label{fig:mutation_testing_process}
\end{figure}

Applying mutation testing to a program involves the following steps: \textit{generation of mutants}, {\em execution of the original program}, {\em execution of mutants}, and {\em analysis of living mutants} to determine whether or not they are equivalent~\cite{offut:2010}.
This process is shown in Figure \ref{fig:mutation_testing_process}, the automated steps are represented in bold boxes, and the manual steps in dashed boxes. 
Given a program $ P $, a set of mutants $ M $ is created from the application of mutation operators in $ P $. Then, a set of test cases $ T $ is created, which can be designed based on the mutants in $ M $ or from another testing criteria based on the choice of the test engineer. This test set $ T $ is then run with the program $ P $. If the tests fail, it is necessary to fix program $ P $. Otherwise, $ T $ is run with the mutants in $ M $. With the test results, the mutation score ($ ms $) is calculated based on the number of dead and alive mutants. From that step, it is necessary to decide whether or not the mutation testing process should continue. This decision can be made based on the value of $ ms $; if it is $ 1.0 $, it means that all non-equivalent mutants have been killed by the test set $ T $. However, it is not always feasible to kill all mutants, given that many mutants can be created. Thus, at the discretion of the test engineer, it is possible to define a threshold value for $ ms $, generally close to $ 1.0 $, so that when this value is reached, the test set $ T $  is considered good, and the process ends.

Mutation testing is one of the most challenging criteria to meet, given the significant effort required to apply all of its steps, as shown in Figure~\ref{fig:mutation_testing_process}.
Given this process and the large number of mutants that can be generated, it is essential that mutation testing be supported by a tool that automates all or part of the process. It is impractical to apply it manually in production.

Over the years, mutation testing has been actively investigated in the area of software testing research~\cite{Jia2011}. Studies have identified mutation testing as one of the most efficient criteria for detecting faults compared to other criteria, such as the works of Franklin et al.~\cite{FRANKL1997235}, Offut et al.~\cite{Offutt:1996:EED:229754.229760} and Walsh~\cite{Walsh:1985:MTC:912107}. These results make mutation testing a reference in the software testing area and are often used as a quality standard to evaluate other criteria, techniques, and test sets in general~\cite{offut:2010}. 
As usual, with very flexible and powerful techniques, mutation testing success depends on how this flexibility is instantiated to each situation.  A good set of mutation operators is needed, leading to the design of tests that explore programs' main potential problems in that specific language or paradigm. Mutation testing for imperative programs requires different mutation operators than those used for a functional language; similarly, an object-oriented program requires some mutation operators to deal with particularities of object-oriented programming. Besides, in the case of frameworks like Apache Spark, Big Data processing code is weaved within a host programming language (e.g., Scala) code. This approach introduces a specific challenge for testing approaches like mutation, which must test ``classic'' code in the host language and propose specific mutation strategies to test the Big Data processing code. Our work addresses mutation testing for the parallel Big Data processing code weaved within Scala programs in the context of Apache Spark like frameworks.

\section{Related Work}\label{sec:related}

Parallel programming models are based on the principle that significant problems can often be divided into smaller ones, which can then be solved simultaneously.  Regarding big data processing, two strategies have emerged, namely  \textit{control flow} and \textit{data flow}-based parallel programming. 
Under the control flow strategy, a single system node controls the entire program execution (master, orchestrator). 
In the data flow parallel model (also called choreography in the Services world), the processes
that execute the program trigger the execution of other program components. 
Lately, many databases
research projects focus on data flow control since it tends to
enhance response times and throughput~\cite{TeeuwBlanken1993}.

The most popular control flow-based model for big data is MapReduce~\cite{Dean2004}. 
A MapReduce program has two main components: A \textsl{map} procedure, which applies the same function to all the elements of a key/value dataset, and a \textsl{reduce} procedure, which aggregates values based on their key. 
In this way, the map component is used to filter or sort data (such as sorting students by the first name into queues, one queue for each name).
The reduce component is usually used to summarize data (like counting the number of students in each queue, yielding name frequencies). 
MapReduce implementations such as Hadoop~\cite{hadoop273} enhance reliability by supporting the construction of fault-tolerant systems.
Since the inner work of these programming systems is expected to be reliable, the weakest link is represented by the procedures supplied by the programmer.

The reliability of programs for processing large volumes of data becomes essential given the large number of computational resources required for this type of program to be executed, which means that failures in programs of this type can generate large losses~\cite{Garg2016}. 
In this context, software testing becomes paramount in developing programs for processing large volumes of data with higher quality and less tendency to fail during production. 
The studies presented by Camargo \textit{et al.}~\cite{Camargo:2013} and Mor\'{a}n \textit{et al.}~\cite{moran2018} show that works on testing in this context have increased interest in recent years and reveal that few systematic techniques and tools have been developed so far and that most have not reached a certain level of maturity. 
Furthermore, most of the existing work has focused on programs that exclusively follow control flow based parallel programming models like \textit{MapReduce}~\cite{moran2018}, showing that the testing of programs for processing large volumes of data is still an open research area. 
In this section, we present the state of the art of functional testing of large data processing programs.


The systematic review presented by Mor\'{a}n \textit{et al.}~\cite{moran2018} identified 54 works related to the validation and verification process of control flow-based programming models, mainly based on the MapReduce model. 
The study revealed that most efforts in the area are focused on program performance, identifying 32 (59\%) works related to performance testing. In contrast, only 12 (22\%) works are focused on functional testing, indicating a potentially promising subject in the area. 
Works in the performance testing domain involve verifying non-functional requirements, such as execution time and computational resources, while functional testing is related to the program's functional behaviour. We focus on functional testing of large-scale data processing programs.
We do not analyze works on performance testing since they are mostly related to benchmarks, which is out of the scope of our work.

We describe the works that test parallel data processing programs verifying their functional behaviour to check whether their functional requirements are met.

\subsection{Testing MapReduce (control flow-based) Programs}

As classic programs, testing MapReduce programs can be done dynamically (\textit{i.e.}, observing programs behaviour at runtime) and statically (\textit{i.e.}, analyzing the source code). 
The following sections describe some dynamic and static testing approaches for MapReduce programs implemented on top of different platforms of the Apache Hadoop ecosystem\footnote{The Apache Hadoop project develops open-source software for reliable, scalable, distributed computing. The Hadoop software library is a framework that allows for the distributed processing of large datasets across clusters of computers using simple programming models (\url{https://hadoop.apache.org}).}.

\paragraph{Dynamic testing of control flow-based programs.}
Dynamic testing consists of the execution of a program to verify its properties.
The work presented by Csallner \textit{et al.}~\cite{Csallner:2011}  focuses on checking the commutativity of the operation reduce in MapReduce programs.  
The authors use a symbolic execution technique to search for faults and generate test cases. 
The technique consists of extracting algebraic expressions from the program representing the conditions that lead the program to different execution paths. 
From this, coded execution paths with correction conditions (which verify commutativity) are derived.
Then, a constraint solver is used to infer input values that violate the correction conditions. 
These values are then converted into test cases for the program.
The authors present an implementation of their technique for MapReduce programs in the Hadoop system. 
A similar approach was applied by Li \textit{et al}~\cite{li2013} to test programs in \textit{Pig Latin}~\cite{Olston:2008}, a \textit{script} language for data processing programs that run on Hadoop. 
The authors expand the technique developed by Csallner~\textit{et al.}~\cite{Csallner:2011} to extract test paths by exploring the different operations of Pig Latin.

Xu \textit{et al.}~\cite{Xu2013} introduce a technique that tests properties in operations of stream processing programs with the \textit{SPL} language~\cite{Hirzel2013}, a language for \textit{stream processing}. 
The authors seek to verify properties that are important for the reliability and optimization of this type of program: non-determinism, selectivity, blocking, statefulness, non-commutativity and partition interference. 
The technique uses random data generation to dynamically check (with execution) whether or not these properties are verified.

In \cite{moran2014} the authors present {\em MRTree}, a hierarchical classification of faults in MapReduce programs. 
The faults presented in the classification are divided into faults related to the \textit{reduce} operation and the \textit{combine} operation (an intermediate reduction operation performed right after \textit{map}).
Some examples of the faults presented in {\em MRTree} are: 
(1) a non-commutative reduce operation that can generate different results for the same dataset if data is processed in different orders; and 
(2) faults related to the key/value data, such as inconsistencies between the key and the value or the issue of an incorrect key/value pair, are checked. 
For each fault, the authors propose {\em ad-hoc} testing directives to mitigate them. 
The same research group defined {\em MRFlow}~\cite{Moran:2015}, a technique that derives a data flow graph from \textit{map} and \textit{reduce} operations. 
From this graph, test cases are generated using \textit{graph-based testing techniques}. 
In both papers \cite{moran2014,Moran:2015} authors do not describe a tool implementing their technique.

The method presented by Mattos~\cite{mattos2011}  uses meta-heuristic search techniques to generate test data dynamically. 
The approach evaluates two algorithms: the genetic and the bacteriological algorithm, and conclude that the latter generates better test data. 
To evaluate the work, the authors applied mutation testing with three mutation operators proposed for MapReduce based programs. They propose an operator that inserts an operation \textit{combine}  with the same behaviour as the \textit{reduce}; an operator that removes the \textit{combine}; and an operator that changes the number of processes that execute the \textit{reduce}. 
These operators simulate semantic faults concerning the understanding of the MapReduce model. 
The conclusion is that the bacteriological algorithm performs well in generating test data and contributes to identifying functional faults, but that it has not contributed to the identification of faults related to the misunderstanding of the MapReduce model.

Li \textit{et al.}~\cite{Li2015} introduce a {\em testing framework} for large data processing programs that extracts test data from real datasets. 
Their goal is to extract a subset of test data from the dataset itself, which will be processed using \textit{input space partitioning techniques}. 
In this way, the \textit{framework} gets a representative subset of data from the original dataset.
This subset of test data is then applied in the validation process of the program under test.

\paragraph{Static testing of control flow (MapReduce) programs.}
Some works test MapReduce programs without considering the program's execution, using formal methods and static analysis. 
A work by Chen \textit{et al.}~\cite{Chen2015} makes use of formal methods to verify the commutativity of \textit{reduce} operations in MapReduce programs. 
Since this program is executed in a parallel and distributed computing environment, aggregation operations must be commutative and associative to ensure a deterministic result. The constraint about determinism is explained by the fact that it is not possible to determine the order in which the values will be processed~\cite{Chen2017}. 
The method presented in~\cite{Chen2015} extracts a series of assertions and properties from the \textit{reduce} operation.
These assertions must be verified through an external program, such as a \textit{model checker}, to attest that the operation is commutative.

A method for static type analysis in MapReduce programs is proposed by D\"orre \textit{et al.}~\cite{dorre2011}. 
The authors aim to identify type incompatibilities in the emission of key/value pairs in \textit{map}, \textit{combine} and \textit{reduce} operations. 
This static analysis is done automatically with the \textsl{SNITCH} tool (\textit{StatIc Type Checking for Hadoop}), proposed by the authors. 
Ono \textit{et al.}~\cite{Ono2011} verify the correction of MapReduce programs.
This correction is done formally using \textit{Coq}~\cite{Bertot2010}, an interactive theorem tester.

\subsection{Data Flow Program Testing}

Data flow programming is a programming paradigm that models a program as a directed graph of the data flowing between operations.
Data flow programming promotes the modelling of programs as a series of connections among operations.
Explicitly defined inputs and outputs connect operations, which work like black boxes. 
Instructions do not impose any constraints on sequencing except for the data dependencies. 
An operation runs as soon as all of its inputs become available. 
Thus, data flow languages are inherently parallel and can work well in large, decentralized systems. 

Data flow programming has been adopted by libraries and environments addressing data analytics and processing like MatLab, R and Simulink.  
Big data parallel processing solutions like Apache Spark~\cite{Zaharia:2010}, DryadLINQ~\cite{yu2008}, Apache Beam~\cite{beam2016} and Apache Flink~\cite{katsifodimos2015apache},  also adopt data flow programming models for implementing programs as  processing data flows. 

In general,  testing a data flow program involves 
\textit{(i)} defining testing criteria, \textit{(ii)} classifying paths on the data flow graph that satisfy these criteria, and \textit{(iii)} developing path predicate expressions to derive test input.

\paragraph{Testing data flow based data analytics programs.}
Data flow data analytics has been promoted by Simulink, Matlab and R environments.  Models implemented as functions to be used within programs in this context can also become complex to test. Therefore, approaches have been devoted to promoting full or partial testing strategies. For example,
the  model checking
engine COVER~\cite{brillout2009mutation} defines a verification methodology
to assess the correctness of Simulink models. 
COVER automatically generates test cases and adopts fault and mutation-based testing.
Therefore, coverage of a Simulink program by a test suite is defined in terms of detecting injected faults. 
The work can compute test suites for given fault models using bounded model checking techniques.
MATmute \cite{movva2015automatic} is based on an approach for automatically generating test suites for Scientific MATLAB code.
The approach introduced by Xu \textit{et al.}~\cite{10.1109/ASE.2013.6693071} improves the dependability of data flow programs
by checking operators for necessary properties. 
The approach is dynamic and involves generating tests whose results are checked
to determine whether specific properties hold or not. 

\paragraph{Testing data flow based big data parallel programs.}
For testing   \textit{Apache Spark}, \textit{DryadLINQ}, \textit{Apache Beam}, and \textit{Apache Flink} programs,  it is possible to implement unit tests using external libraries and  functions provided by these systems. 
Libraries such as those defined in~\cite{karau2015b} and~\cite{ottogroup2016} have adopted this strategy. They offer several utility classes for unit testing in Apache Spark and Apache Flink, respectively. 
For Apache Beam and DryadLINQ, it is possible to use native support that enables the definition of program entries, automates the execution of tests and proposes \textit{test oracle} to verifying program results. 
Although these libraries are essential for implementing and executing tests, they do not support test cases' design, which is a critical part of the testing process.

Regarding test case design, the work by Riesco \textit{et al.}~\cite{riesco2015} applies  {\em property-based testing}~\cite{Hughes2000} to test Spark programs. 
This technique consists of generating random data, running the program under test with this input data, and then verifying the program's behavior through oracles that verify if the program's results meet specific defined properties through logical expressions. 
In \cite{riesco2015} the authors presented \textit{sscheck}, a library for the application of property-based testing in Spark programs. 
That work was extended in~\cite{riesco2016} and~\cite{riesco2019} to test stream processing programs with {\em Spark Streaming}. 
The authors applied temporal logic to verify time-related properties, which is an essential variable in stream processing. 
Their work was adapted for Apache Flink in~\cite{Espinosa2019}.

\subsection{Discussion}

The exploitation of parallelism by data processing programs is different from the use of parallelism in a general
purpose computer system. 
Data processing programs operate on
large amounts of data, making the data distribution across the system nodes significantly important.
In classic parallel processing programs system, parallelism is vital for avoiding bottlenecks. 
In contrast, in a data processing program, large streams of input and output data make the network and the disk I/O the bottleneck rather than CPU usage. 
Consequently, the use of parallelism is strongly
determined by the communication across the nodes of the
system. 
Also, parallelism will be exploited to enhance parallel
disk I/O. 
A consensus on parallel and distributed data processing system
architecture has emerged, based on a so-called \textit{shared
nothing} hardware design. 
In a shared
nothing system several nodes, each having its own
processor, memory modules and secondary storage devices,
are connected by a local area network (LAN). 
The only
way processors communicate with one another is by sending
messages via this interconnection network. 

Parallel programs do not only include data processing operations (filtering, selection, clustering); they include data I/O directives used to guide control and data sharing across nodes.  
Testing a control and data flow of a parallel processing program implies testing how data processing operations and data management operations are coordinated during the execution of a parallel program. 
In both cases, the code implementing the data processing operations using a target programming language does not introduce further challenges than classic programs testing. 
However, program testing must include the code used to deal with read/write operations, global/local variables used to share data in memory/cache, and exchange it across processes.  
In the case of data flow based programs, testing must consider the way data management operations are weaved in the program logic.

The works presented previously focusing on parallel programs using the MapReduce programming model do not provide tools for automatically testing programs, or they have not been widely experimented~\cite{Camargo:2013, moran2018}. 
The principle of testing data flow-based programs involves determining some programs' properties by analyzing or verifying properties on the data flow. 
In general, the challenge consists in (automatically) generating tests and then developing tools for evaluating these tests on top of programs. 
The characteristics of the programming languages promoting a data flow programming model guide the type of generated tests that consider input and output data of operations applies on top of them.
This absence of testing support shows that testing large volume data processing programs still lacks techniques and tools, mainly for testing programs that follow a data flow model.

\section{Mutation Operators for Apache Spark Big Data Processing programs}\label{sec:mutation_operators}

Mutation testing is based on the notion of ``mutation operator'' that defines the rules that state how to add changes to a program to simulate faults. Therefore, the faults in a program are, in general, modelled beforehand according to the characteristics of the programming language. For example, faults can be related to a missing iteration in a loop or a mistake in an arithmetical or logical expression.

We have proposed a set of mutation operators for Spark programs~\cite{caise2020}. These mutation operators mimic common faults and mistakes in Spark programs that we identified in a previous study and classified into a taxonomy~\cite{joaothesis}.  These faults concern the incorrect definition of a program's data flow, like calling transformations in a wrong order;  and introducing mistakes when calling transformations (e.g., calling the wrong transformation or passing the wrong parameter in a transformation call).

In the following lines, we briefly describe the taxonomy~\cite{joaothesis} that classifies a set of functional faults that concern  Apache Spark programs. This taxonomy was used as a reference to design the mutation operators that we proposed~\cite{caise2020} (we refer to this approach as \textit{transformation mutation}). This set of mutation operators specify modifications on:
 \begin{enumerate}
     \item  the DAG that defines the program's \textit{data flow}, for example, changing the calling order of two transformations;
     \item specific \textit{transformations} (see Section \ref{sec:background}), such as changing the function passed as input to an aggregation transformation. 
 \end{enumerate}
 The interested readers can refer to the paper \cite{sbmf2020} that describes these operators' formalization and their application to data flow parallel Big Data processing programs in popular engines like DryadLINQ~\cite{yu2008}, Apache Beam~\cite{beam2016} and Apache Flink~\cite{katsifodimos2015apache} that implement a data flow model similar to Apache Spark.

\subsection{Functional Faults in Apache Spark Programs}

The design of big data processing code within Spark programs includes making sure to avoid functional faults that correspond to	inconsistencies within the code that may result in unexpected behaviour when executed. Examples of these faults include wrong transformations, incorrect parameters and absence of transformations.

\begin{figure}
	\centering
	\includegraphics[width=.6\textwidth]{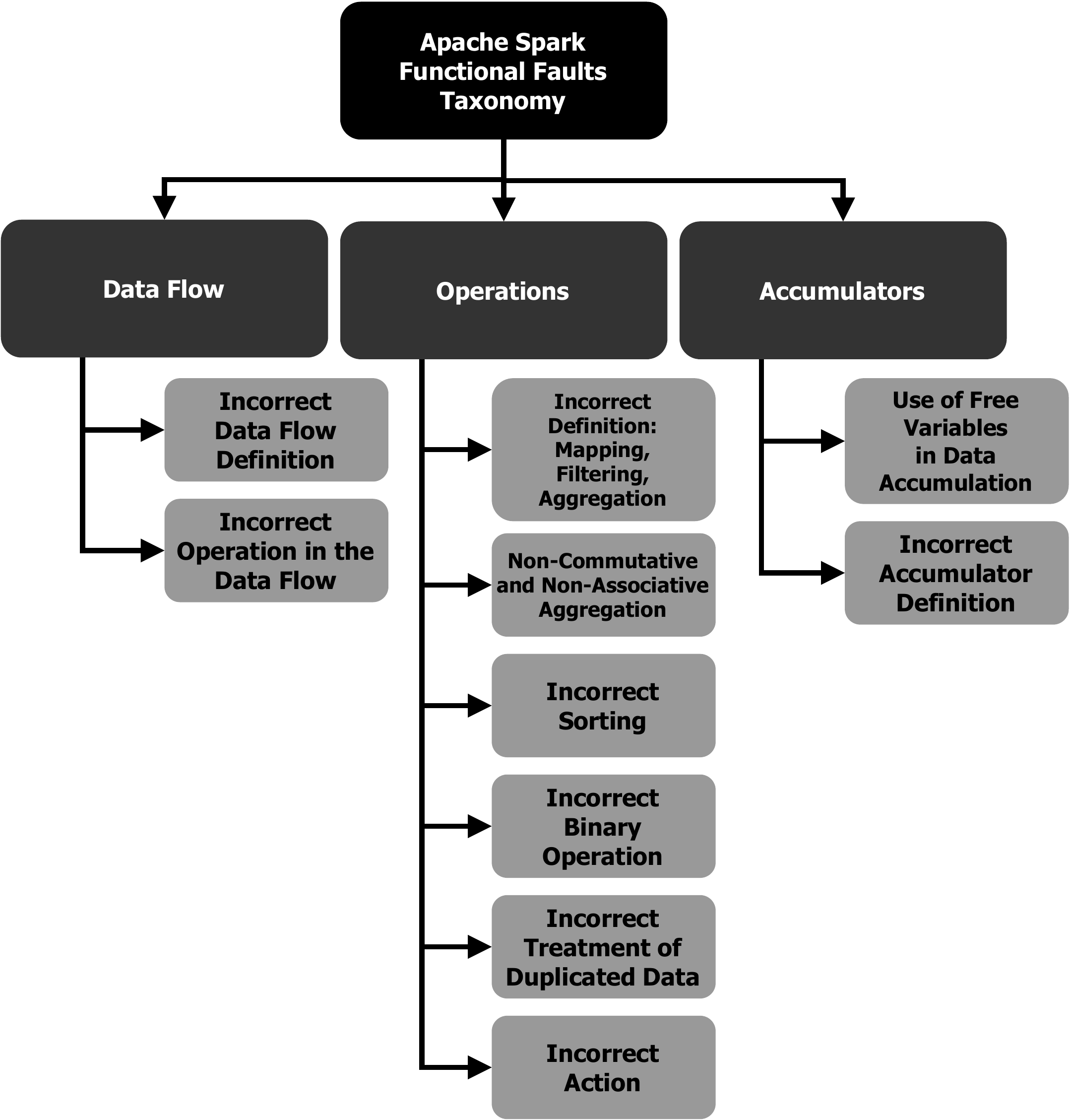}
	\caption{Taxonomy of Apache Spark Functional Faults~\cite{joaothesis}.}
	\label{fig:spark-functional-faults-taxonomy}
\end{figure}

The functional faults that can come up in Apache Spark Big Data processing programs are classified in the taxonomy proposed by Souza Neto~\cite{joaothesis} and shown in Figure~\ref{fig:spark-functional-faults-taxonomy}. The taxonomy was proposed based on an analysis of Apache Spark documentation, other references in the literature and an analysis of the source codes of a set of Spark programs.
Through this study, we identified a set of faults that we organized in a taxonomy. The taxonomy classifies three types of functional faults related to the data flow, the strategy and order in which operations are called, and the use of accumulators.

Faults related to the \textit{data flow} of a Spark program refer to (i) an incorrect operations' call order; and (ii) using the wrong data processing operation.
To illustrate the first case, let us consider the Spark code shown in Figure~\ref{fig:analise-log-spark} that analyzes log files to count the number of error messages happening in the log. Under the hypothesis that error messages contain the string ``ERROR'' on the top, the code implementing a correct version of this task is shown in Figure~\ref{fig:analise-log-spark-a}. The code first filters the messages starting with the word ``ERROR'' at the top; then, it crops the beginning of each message applying a map function. Finally, it filters messages containing the word  ``foo''.

In contrast, Figure~\ref{fig:analise-log-spark-b} shows a version of the code with the following faults in the data flow.   The two filtering operations are called in the wrong order  (inversion of lines 2 and 4) tagged with $ \Delta_1 $ and $ \Delta_3 $. 
By inverting the filters, the result is entirely different. First, the program obtains a list of messages containing the word ``foo'', thus messages containing ``ERROR'' on the top and other messages. Then, the function map crops the messages initial substring. Consequently, the substring ``ERROR'' is deleted from messages that correspond to those intended to be retrieved. The final result is a list of messages that do not necessarily contain ``ERROR'' at the top or contain this string in their ``body''.

The line tagged with $ \Delta_2 $ in the code illustrates the incorrect use of an operation in a data flow. The program statement uses the  wrong type of map function since it calls \textit{flatMap} instead of \textit{map}. 
The operation \textit{map}  expresses a one-to-one transformation, transforming each element of a collection into one element of the resulting collection. In contrast, the operation \textit{flatMap} expresses a one-to-many transformation, so it transforms each element of a collection to 0 or more elements of the resulting collection. Regarding the \textit{map} operation applied on line 3 (see Figure~\ref{fig:analise-log-spark-a}) it splits the log message based on the separator ``\lstinline[style=Scala]|\\t|'' and filters  the third element from the resulting data (RDD). The second version of the code (Figure~\ref{fig:analise-log-spark-b}) the operation \textit{flatMap}  (line 3) only splits the log message based on the separator ``\lstinline[style=Scala]|\\t|''. The resulting data (RDD)  contains all the strings containing this separation. Nothing is done with the third element in comparison to the first version. Both programs produce a result of the same data type  (\lstinline[style=Scala]|RDD[String]|), but with  different content.
    
The second type of fault identified in the taxonomy (see Figure \ref{fig:spark-functional-faults-taxonomy}) concern the way \textit{operations} are used and weaved to process data in a program. These faults include the incorrect definition of mapping, filtering and aggregation operations, like passing an incorrect processing function as a parameter to the transformation. For instance, consider the examples of faults shown in Figure~\ref{fig:analise-log-spark-c}. An example of an incorrect definition of filtering transformations in the code is tagged with $\Delta_4$ and $\Delta_6$. 
The line tagged with $\Delta_4$ verifies whether the message string bottom corresponds to the word ``ERROR'' instead of looking for that string in the top. The line tagged with $\Delta_6$ checks whether the whole message corresponds to the string ``foo'' instead of just checking if it contains that word.
The line tagged with $\Delta_5$ illustrates the incorrect definition of a mapping transformation with erroneous functions passed as parameters.  
In this case, the separator (`` '') passed to the function \textit{split}   called within the function \textit{map} is different from the separator used in the correct version (``\lstinline[style=Scala]|\\t|''). Thus, the subset of strings that will be generated by splitting the message with `` '' will be different from those generated  using ``\lstinline[style=Scala]|\\t|''. Thus, the   selection in the mapping will produce different results in both programs since the function selects  the third substring.

\begin{figure}
\centering
\begin{subfigure}[t]{0.4\textwidth}
\begin{CenteredBox}
\begin{lstlisting}[style=Scala, mathescape=true]
val fooCount = logsRDD
	.filter(_.startsWith("ERROR"))
	.map(_.split('\t')(2))
	.filter(_.contains("foo"))
	.count	
\end{lstlisting}
\end{CenteredBox}
\caption{Correct Program}
\label{fig:analise-log-spark-a}
\end{subfigure}%
~
\begin{subfigure}[t]{0.4\textwidth}
\begin{CenteredBox}
\begin{lstlisting}[style=Scala, mathescape=true]
val fooCount = logsRDD
$\Delta_1$	.filter(_.contains("foo"))
$\Delta_2$	.flatMap(_.split('\t'))
$\Delta_3$	.filter(_.startsWith("ERROR"))
		.count	
\end{lstlisting}
\end{CenteredBox}
\caption{Faults in the Data Flow}
\label{fig:analise-log-spark-b}
\end{subfigure} \\[4ex]
\begin{subfigure}[t]{0.4\textwidth}
\begin{CenteredBox}
\begin{lstlisting}[style=Scala, mathescape=true]
val fooCount = logsRDD
$\Delta_4$	.filter(_.endsWith("ERROR"))
$\Delta_5$	.map(_.split(' ')(2))
$\Delta_6$	.filter(_.equals("foo"))
		.count	
\end{lstlisting} 
\end{CenteredBox}
\caption{Faults in Operations}
\label{fig:analise-log-spark-c}
\end{subfigure}\\[4ex]
\caption{Example of a Spark program for log analysis.}%
\label{fig:analise-log-spark}
\bigskip
\end{figure}

This type of faults can also come up when using the following operations: (i) Choosing incorrect binary operations (joins and set-like operations), for example calling the wrong type of join operation\footnote{Note that Spark proposes different implementations of the operator join with different semantics. The programmer must be sure of the type of join pertinent to the application logic.}; (ii) Choosing the wrong sorting operations, for example, choosing the wrong order option (ascending or descending); (iii) Incorrect treatment of duplicate data, like not removing duplicate data from a result when necessary.

The third group of faults in the taxonomy (see Figure \ref{fig:spark-functional-faults-taxonomy}) concern \textit{Accumulators}. \textit{Accumulators} are variables that are only ``added'' through an associative operation and can, therefore, be efficiently supported in parallel. They can be used to implement counters (as in MapReduce) or sums~\cite{spark22}. Spark natively supports accumulators of numeric types, and programmers can add support for new types. 
In the example below, we create an accumulator variable (\textit{accum}) and use it to aggregate values within a Spark operation:
    
\begin{CenteredBox}
\begin{lstlisting}[style=Scala, numbers=none]
var accum: LongAccumulator = sc.longAccumulator
sc.parallelize(Array(1, 2, 3, 4)).foreach(x => accum.add(x))
\end{lstlisting}
\end{CenteredBox}
    

Faults  related to accumulators concern either their commutative and associative properties and their use. Indeed, accumulators in Spark can be commutative and associative and the right type must be chosen in the code depending on the way data should be shared across processes to aggregate partial results. The implementation of some tasks, require absolutely the use of accumulators to produce the right result. The programmer must have a clear understanding of the properties of the task to know that its implementation requires the use of accumulators.
The example below illustrates a mistake made when using a standard variable instead of an accumulator. Spark handles variables and accumulators differently, a variable will be copied and managed locally on each node in the cluster, so the final result will not contain the total sum produced in parallel by all the nodes. In contrast with an accumulator, the final result will aggregate the partial sums computed by each node.

    \begin{CenteredBox}
    \begin{lstlisting}[style=Scala, numbers=none]
    var accum: Long = 0
    sc.parallelize(Array(1, 2, 3, 4)).foreach(x => accum += x)
    \end{lstlisting}
    \end{CenteredBox}



Through our study, we identified 16 possible faults that can come up in Spark programs. Refer to~\cite{joaothesis} for details on this study. Next we define the mutation operators that we propose based in these faults~\cite{caise2020}, namely mutation operators for:  \textit{data flow} and  \textit{transformations}.

 \subsection{Mutation Operators for Data Flow}

The operators defined in this group replace, swap and delete unary and binary transformations calls. Recall that a unary transformation operates on a single input RDD and a binary transformation on two input RDDs.

\smallskip
\noindent\textit{Unary Transformations Replacement (UTR)} - 
replaces a unary transformation for another with the same input and output signature (i.e., both transformations receive and return RDDs of the same type).

\smallskip
\noindent\textit{Unary Transformation Swap (UTS)} - 
swaps the calling order of two unary transformations in the program, provided that they have the same input and output signature.

\smallskip
\noindent\textit{Unary Transformation Deletion (UTD)} - 
removes the call of a unary transformation that receives and returns RDDs (both) of the same type in the program.
\smallskip

The operators \textit{Binary Transformation Swap} (BTS) and \textit{Binary Transformations Replacement} (BTR) operate on binary transformations.

Operators in this group change the DAG that defines the data flow of a Spark program (see the group data flow in the taxonomy in Figure~\ref{fig:spark-functional-faults-taxonomy}). 

To illustrate the mutation operators in this group, let us consider the program shown in Figure~\ref{fig:analise-log-spark-a} that filters error messages in a log. Table~\ref{tab:mutants-data-flow-example} shows examples of the mutants that can be generated for this program applying the data flow mutation operators. Mutant 1 was generated by the operator UTR, replacing the filtering transformation in line 2 with the mapping transformation in line 3 and no changes in lines 3 and 4. 
Mutant 2 was generated by the operator UTS  swapping lines 2 and 3. Mutant 3 was generated by the operator UTD deleting the filtering transformation from line 4.

\begin{table}[htbp!]
\bigskip
\centering
\caption{Examples of mutants generated with the data flow mutation operators.}
\label{tab:mutants-data-flow-example}
\begin{tabular}{|c|c|c|l|}
\hline
\textbf{Id} & \textbf{Operator} & \textbf{Lines} & \multicolumn{1}{c|}{\textbf{Mutation}} \\ \hline
1 & UTR & 2 & 
\begin{lstlisting}[style=Scala, numbers=none]
val fooCount = logsRDD
   .map(_.split('\t')(2))
   .map(_.split('\t')(2))
   .filter(_.contains("foo"))
   .count
\end{lstlisting}  \\ \hline
2 & UTS & 2, 3 & 
\begin{lstlisting}[style=Scala, numbers=none]
val fooCount = logsRDD
   .map(_.split('\t')(2))
   .filter(_.startsWith("ERROR"))
   .filter(_.contains("foo"))
   .count
\end{lstlisting}  \\ \hline
3 & UTD & 4 & 
\begin{lstlisting}[style=Scala, numbers=none]
val fooCount = logsRDD
   .filter(_.startsWith("ERROR"))
	  .map(_.split('\t')(2))
	  .count
\end{lstlisting}  \\ \hline
\end{tabular}
\end{table}

\subsection{Mutation Operators for Transformations}

The operators of this group replace, invert, insert and delete specific transformations in a Spark program: mapping,  filtering, set-like, distinct, aggregation, join and order.   Table~\ref{tab:mutants-example} shows examples of the mutation operators of this group using excerpts from the Spark programs shown in Figure~\ref{fig:wordcount-example} and Figure~\ref{fig:analise-log-spark-a}. We also use other code snippets to illustrate mutation operators that modify transformations that are not applied in previous programs' examples. We also define reductions rules that define the conditions in which the application of some operators override the application of others.

\begin{table}
\caption{Examples of mutants generated with the transformation mutation operators.}
\label{tab:mutants-example}
\begin{adjustbox}{width=1\textwidth}
\begin{tabular}{|l|l|l|l|l|}
\hline
\textbf{Id} & \textbf{Operator} & \textbf{Fig.} & \textbf{Line} & \textbf{Mutation}                                                                              \\ \hline
1  & MTR      & \ref{fig:wordcount-example} & 2    & \lstinline[style=Scala]|val words = input.flatMap( (line: String) => ...originalValue.headOption )| \\ \hline
2  & MTR      & \ref{fig:wordcount-example} & 2    & \lstinline[style=Scala]|val words = input.flatMap( (line: String) => ...originalValue.tail )|       \\ \hline
3  & MTR      & \ref{fig:wordcount-example} & 2    & \lstinline[style=Scala]|val words = input.flatMap( (line: String) => ...originalValue.reverse )|    \\ \hline
4  & MTR      & \ref{fig:wordcount-example} & 2    & \lstinline[style=Scala]|val words = input.flatMap( (line: String) => ...Nil )|                      \\ \hline
5  & NFTP      & \ref{fig:analise-log-spark-a} & 4    & \lstinline[style=Scala]|...   .filter(!_.contains("foo"))|                      \\ \hline
6  & STR      & -- & --    & \lstinline[style=Scala]|val rdd3 = rdd1.union(rdd2)|                      \\ \hline
7  & STR      & -- & --    & \lstinline[style=Scala]|val rdd3 = rdd1.intersection(rdd2)|                      \\ \hline
8  & STR      & -- & --    & \lstinline[style=Scala]|val rdd3 = rdd1|                      \\ \hline
9  & STR      & -- & --    & \lstinline[style=Scala]|val rdd3 = rdd2|                      \\ \hline
10  & STR      & -- & --    & \lstinline[style=Scala]|val rdd3 = rdd2.subtract(rdd1)|                      \\ \hline
11  & DTD      & -- & --    & \lstinline[style=Scala]|val rdd4 = rdd3|                      \\ \hline
12 & DTI      & \ref{fig:wordcount-example} & 2    & \lstinline[style=Scala]|val words = input.flatMap( (line: String) => line.split(" ") ).distinct()|  \\ \hline
13 & DTI      & \ref{fig:wordcount-example} & 3    & \lstinline[style=Scala]|val pairs = words.map( (word: String) => (word, 1) ).distinct()|            \\ \hline
14 & DTI      & \ref{fig:wordcount-example} & 4    & \lstinline[style=Scala]|val counts = pairs.reduceByKey( (a: Int, b: Int) => a + b) ).distinct()|    \\ \hline
15 & ATR      & \ref{fig:wordcount-example} & 4    & \lstinline[style=Scala]|val counts = pairs.reduceByKey( (a: Int, b: Int) => a) )|                   \\ \hline
16 & ATR      & \ref{fig:wordcount-example} & 4    & \lstinline[style=Scala]|val counts = pairs.reduceByKey( (a: Int, b: Int) => b) )|                   \\ \hline
17 & ATR      & \ref{fig:wordcount-example} & 4    & \lstinline[style=Scala]|val counts = pairs.reduceByKey( (a: Int, b: Int) => a + a )|               \\ \hline
18 & ATR      & \ref{fig:wordcount-example} & 4    & \lstinline[style=Scala]|val counts = pairs.reduceByKey( (a: Int, b: Int) => b + b )|               \\ \hline
19 & ATR      & \ref{fig:wordcount-example} & 4    & \lstinline[style=Scala]|val counts = pairs.reduceByKey( (a: Int, b: Int) => b + a )|               \\ \hline
20 & JTR & -- & -- &  
\begin{lstlisting}[style=Scala, numbers=none]
val rdd4 = rdd3.leftOuterJoin(rdd2)
                .map(x => (x._1, 
                 (x._2._1, x._2._2.getOrElse(""))))
\end{lstlisting}\\ \hline
21 & JTR & -- & -- &  
\begin{lstlisting}[style=Scala, numbers=none]
val rdd4 = rdd3.rightOuterJoin(rdd2)
                .map(x => (x._1, 
                 (x._2._1.getOrElse(0), x._2._2)))
\end{lstlisting}\\ \hline
22 & JTR & -- & -- &  
\begin{lstlisting}[style=Scala, numbers=none]
val rdd4 = rdd3.fullOuterJoin(rdd2)
                .map(x => (x._1, 
                 (x._2._1.getOrElse(0), x._2._2.getOrElse(""))))
\end{lstlisting}\\ \hline
23 & OTD      & -- & --    & \lstinline[style=Scala]|val rdd2 = rdd1|               \\ \hline
24 & OTI      & -- & --    & \lstinline[style=Scala]|val rdd2 = rdd1.sortByKey(ascending = false)|               \\ \hline
\end{tabular}
\end{adjustbox}
\end{table}

\smallskip
\noindent\textit{Mapping Transformation Replacement (MTR)} - 
Given a mapping transformation (\textit{map}, \textit{flatMap}) that receives a  mapping function $f$  as  parameter, the operator MTR replaces $f$ by a mapping function $f_m$, where (a) $f_m$ returns a constant value of the same type returned by $f$, or (b) $f_m$ modifies the value returned by $f$. For example, for a mapping function that operates on a parameter of type Integer, the operator MTR defines five cases where $f_m$ defines the following value types to be returned:  the constants $0$, $1$, $Max$ (denoting the highest value of the integer type)  
and $Min$ (denoting the lowest value of the integer type) and the original output value of $f$ but with an inverted sign.

Table~\ref{tab:mapping_values} shows mapping values of basic types and collections that we defined as output parameter types for $f_m$. In the table, $x$ represents the value generated by the original mapping function; $k$ and $v$ represent the key and value generated by the original mapping function in the case of key/value tuples; $k_m$ and $v_m$ represent modified values for the key and value, that correspond to the application of other mapping values respecting the type.

\begin{table}[!htbp]
    \bigskip
	\centering
	\caption{Mapping values for basic and collections types.}
	\label{tab:mapping_values}
	\begin{tabular}{|c|c|}
		\hline 
		\textbf{Type}      & \textbf{Mapping Value}                   \\ \hline
		Numeric & $ 0, 1, MAX, MIN, -x  $                   \\ \hline
		Boolean   & $ true, false, \neg x $                      \\ \hline
		String     & `` ''                         \\ \hline
		List  & $List(x.head)$, $x.tail$, $x.reverse$, $Nil$ \\ \hline    
		Tuple  & $(k_m, v)$ , $(k, v_m)$ \\ \hline    
		General & \textit{null} \\ \hline
	\end{tabular} 
\end{table}

To illustrate the MTR operator, consider the mapping transformation applied in line 2 of the words count program shown in Figure~\ref{fig:wordcount-example} (see this excerpt below). The MTR operator generates mutants 1--4 in Table~\ref{tab:mutants-example}. For the sake of simplicity, we are only showing the modified line, hiding some details, such as calling the original function to obtain the original value we need in some mutants (\textit{originalValue} refers to this original value).

\begin{CenteredBox}
\begin{lstlisting}[style=Scala, numbers=none]
val words = input.flatMap( (line: String) => line.split(" ") )
\end{lstlisting}
\end{CenteredBox}

\smallskip
\noindent\textit{Filter Transformation Deletion (FTD)} - 
 this operator creates a mutant for  each \textit{filter} transformation call in a given program, deleting one \textit{filter} at a time.

{\em Reduction rule:} This operator is a specific case of UTD (unary transformation deletion) because it generates mutants where filters that are a type of unary transformations have been deleted.  So during a mutation process, FTD is applied if UTD has not been applied before.

Mutant 3, shown in Table~\ref{tab:mutants-data-flow-example} illustrates mutant 3 generated with the operator UTD.  The operator FTD applied to the filtering transformation in line 4 of Figure~\ref{fig:analise-log-spark-a} generates this mutant. Note in the excerpt below that it has been removed from the program.

\begin{CenteredBox}
\begin{lstlisting}[style=Scala, numbers=none]
...   .filter(_.contains("foo"))
\end{lstlisting}
\end{CenteredBox}

\smallskip
\noindent\textit{Negation of Filter Transformation Predicate (NFTP)} - 
Given a \textit{filter} transformation call with a predicate function $p$ given as input parameter, the operator NFTP replaces the predicate function $p$ with a predicate function $p_m$ that negates the result of the original function ($p_m(x) = \neg p(x)$).
As an example of this operator, consider the same filtering transformation used to illustrate the operator FTD. The operator NFTP  in this transformation generates mutant 5 in Table~\ref{tab:mutants-example}.

{\em Reduction rule:} The application of this operator is overridden by applying the operators UTD or FTD. Even if the mutants generated by the operator NFTP are not generated by the operators UTD or FTD, we experimentally observed that tests designed to kill the mutants that delete unary/filtering transformations kill the mutants that negated the filtering predicate (as those generated by NFTP).

\smallskip
\noindent\textit{Set Transformation Replacement (STR)} - 
for each occurrence of a set-like transformation (\textit{union}, \textit{intersection} and \textit{subtract}) in a program, this operator creates five mutants: (1-2) replacing the transformation by each of the other remaining set transformations, (3) keeping just the first RDD, (4) keeping just the second RDD, and (5) changing the order of the RDDs in the transformation call (only for the \textit{subtract} transformation since \textit{union} and \textit{intersection} are commutative). 
For example, given the following excerpt of code with a \textit{subtract} between two RDDs:
\vspace{2mm}
\begin{CenteredBox}
\begin{lstlisting}[style=Scala, numbers=none]
val rdd3 = rdd1.subtract(rdd2)
\end{lstlisting}
\end{CenteredBox}

The operator STR  applied on this transformation creates the five mutants, described at lines 6--10 in Table~\ref{tab:mutants-example}.

\smallskip
\noindent\textit{Distinct Transformation Deletion (DTD)} - 
for each call of a \textit{distinct} transformation in the program, this operator creates a mutant by deleting its call. As the \textit{distinct} transformation removes duplicated data from the RDD, this mutation keeps the duplicates. 
For example, the application of DTD in the following excerpt of code generates the  mutant 11 of Table~\ref{tab:mutants-example}:

\begin{CenteredBox}
\begin{lstlisting}[style=Scala, numbers=none]
val rdd4 = rdd3.distinct()
\end{lstlisting}     
\end{CenteredBox}

{\em Reduction rule:} The DTD operator is a specific case of the operator UTD. It generates mutants that are also generated by the operator UTD. So  DTD is applied if UTD has not been applied before.

\smallskip
\noindent\textit{Distinct Transformation Insertion (DTI)} - 
for each transformation (other than \textit{distinct}) in the program, this operator creates a mutant inserting a \textit{distinct} transformation call after that transformation. 
Applying DTI to the transformations presented in Figure~\ref{fig:wordcount-example} generates the mutants 12--14 of Table~\ref{tab:mutants-example}.

\smallskip
\noindent\textit{Aggregation Transformation Replacement (ATR)} - 
Given an aggregation transformation with an aggregation function $f$  as input parameter
the operator ATR replaces $f$ by a different aggregation function $f_m$. 
The definition of ATR considers five replacement functions. Given an original function $f(x,y)$, the corresponding replacement functions $f_m(x,y)$ are: (1) a function that returns the first parameter ($ f_m(x,y) = x $); (2) a function that returns the second parameter ($ f_{m}(x,y) = y $); (3) a function that ignores the second parameter and calls the original function with a duplicated first parameter ($ f_{m}(x,y) = f(x,x) $); (4) a function that ignores the first parameter and calls the original function with a duplicated second parameter ($ f_{m}(x,y) = f(y,y) $); and (5) a function that swaps the order of the parameters ($ f_{m}(x,y) = f(y,x) $), which generates a different value for non-commutative functions. 
Table~\ref{tab:mutants-example} shows the mutants 15--19 as examples of mutants generated by the operator ATR applied to the aggregation transformation on line 4 in Figure~\ref{fig:wordcount-example} (see this excerpt below).

\begin{CenteredBox}
\begin{lstlisting}[style=Scala, numbers=none]
val counts = pairs.reduceByKey( (a: Int, b: Int) => a + b) )
\end{lstlisting}     
\end{CenteredBox}

\smallskip
\noindent\textit{Join Transformation Replacement (JTR)} - 
for each occurrence of a join transformation ((inner) \textit{join}, \textit{leftOuterJoin}, \textit{rightOuterJoin} and \textit{fullOuterJoin}) in a program, the operator JTR replaces that transformation  by the three remaining join transformations. Additionally, a map transformation is inserted after the new join to adjust it, typing with the replaced one. This adjustment is necessary to maintain the type consistency between the mutant and the original program. Indeed depending on the join type, the left, right, or both sides can be optional, and the resulting RDD can be slightly different. 
For example, replacing the transformation (inner) \textit{join}  by \textit{rightOuterJoin} makes left-side values optional. To keep type consistency with the original transformation, we map empty left-side values to default values, in case of basic types, or \textit{null}, otherwise.

To illustrate the operator JTR, let us consider the following code snippet where two RDDs are joined. Assume that \textit{rdd3} is of type \lstinline[style=Scala]|RDD[(Int, Int)]| and that \textit{rdd2} is of type \lstinline[style=Scala]|RDD[(Int, String)]|. The resulting RDD of this join (\textit{rdd4}) is of type \lstinline[style=Scala]|RDD[(Int, (Int, String))]|. Applying JTR to this transformation generates the mutants 20--22 of Table~\ref{tab:mutants-example}. Taking mutant 21 as an example, replacing \textit{join} with \textit{rightOuterJoin}, the resulting RDD is of type \lstinline[style=Scala]|RDD[(Int, (Option[Int], String))]|. Thus, the \textit{map} following the \textit{rightOuterJoin} serves to set the value of type \lstinline[style=Scala]|Option[Int]| to \lstinline[style=Scala]|Int|. When this value is empty (\textit{None}), we assign the value zero (\lstinline[style=Scala]|0|).

\begin{CenteredBox}
\begin{lstlisting}[style=Scala, numbers=none]
val rdd4 = rdd3.join(rdd2)
\end{lstlisting}     
\end{CenteredBox}

\smallskip
\noindent\textit{Order Transformation Deletion (OTD)} - 
for each order transformation (\textit{sortBy} and \textit{sortByKey}) in the program, this operator creates a mutant where the call to that transformation is deleted from the program.
Considering the code snippet below, the operator OTD  generates mutant 23 from Table~\ref{tab:mutants-example}.

\begin{CenteredBox}
\begin{lstlisting}[style=Scala, numbers=none]
val rdd2 = rdd1.sortByKey()
\end{lstlisting}     
\end{CenteredBox}

{\em Reduction rule:} The operator OTD is a specific case of UTD that generates mutants that delete unary transformations of a specific type (order transformations). These transformations are also generated with the operator UTD that considers any unary transformation. So the application of UTD overrides the application of OTD.

\smallskip
\noindent\textit{Order Transformation Inversion (OTI)} - 
for each order transformation in a program, the operator OTI creates a mutant where the ordering (ascending or descending) is replaced by the inverse one. Considering the same code snippet that was used as an example for the OTD operator, the application of the OTI operator in this transformation generates mutant 22 in Table~\ref{tab:mutants-example}, where the ascending ordering that is true by default was changed for false.

{\em Reduction rule:}
The operators UTD and OTD override the operator OTI. Even if the operator OTI generates different mutants as those generated by UTD and OTD, we experimentally observed tests designed to kill the mutants that delete an order transformation   (as those generated with UTD or OTD)  also kill the mutants generated by OTI that invert the order.

\section{\transmut}\label{sec:transmut}

This section presents \transmut{} (\textit{Transformation Mutation for Apache Spark}), a tool that automates the mutation testing process of Spark programs written in Scala. 
The tool was developed as a \textit{plugin} for \textit{SBT} (\textit{Scala Build Tool})~\cite{Suereth2015}, a tool for building projects in Scala and Java. 
\transmut{} automates the process of generating the mutants, applying the mutation operators presented in Section~\ref{sec:mutation_operators}, executing the tests on the original program and the mutants, and analyzing the test results, generating a report with metrics and process results.

\subsection{Functionalities}\label{sec:transmut-funcionalidades}
\transmut{} automates the main steps of the mutation testing process, which includes the processes of mutants generation and execution of the tests with the mutants. The main functionalities implemented by the tool are:

\paragraph*{Program analysis:} 

\transmut{} receives as input a source code containing the Spark program under test. This code is analyzed to identify the principal elements and places  necessary to apply the mutation operators, which include the datasets (RDDs) and the transformations applied in the program, as well as their data flow (DAG);

\paragraph*{Mutant generation:} 

\transmut{} generates mutants for the Spark program under test by applying the mutation operators presented in Section~\ref{sec:mutation_operators}.
The generated mutants are incorporated in a single source code, called \textit{meta-mutant}, in order to reduce the amount of code to be compiled and managed;
	
\paragraph*{Test execution:} 
\transmut{} starts by executing the tests with the original program and checks if their results are as expected.
If so, the tool executes the tests with each mutant and stores their results for analysis;
	
\paragraph*{Mutant analysis:} 
\transmut{} analyzes the results of the tests to verify which mutants were killed by the test set. 
After this, the tool calculates the mutation score and other metrics to generate a report.

\subsection{Mutation Process Workflow}

\begin{figure}[!hbtp]
	\centering
	\includegraphics[width=.95\textwidth]{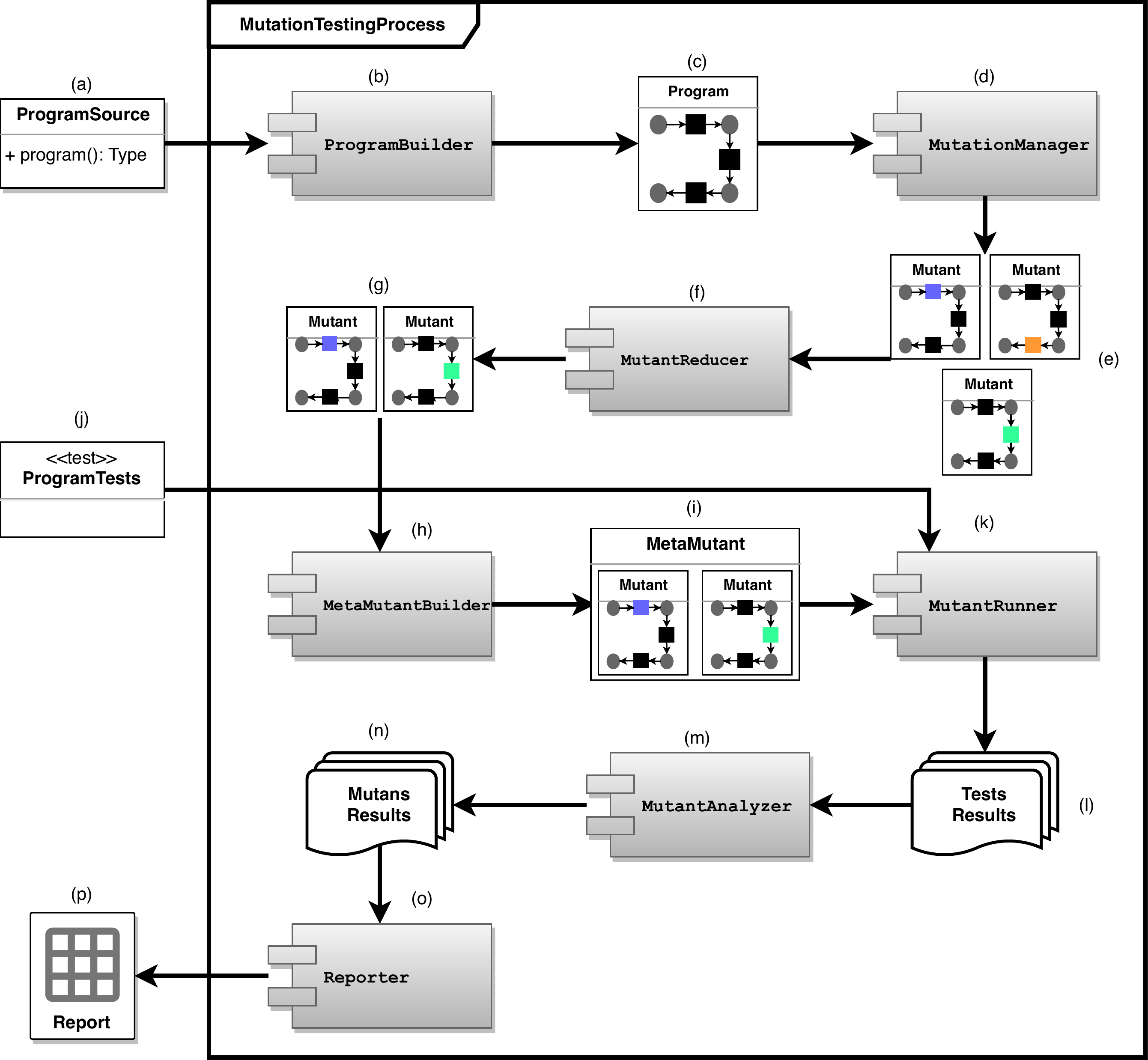}
	\caption{Overview of the workflow of \transmut.}
	\label{fig:transmut-process}
\end{figure}

Figure~\ref{fig:transmut-process} shows an overview of the workflow implemented by \transmut{} to perform the mutation testing process in Spark programs. The figure shows the tool's main modules with their input and output and the flow among modules. The process and the modules are detailed below.

The tool receives a source code as input containing a method that implements the Spark program \textit{(a)}. This code is passed as input to the \chl{ProgramBuilder} module \textit{(b)} which performs a syntactic and type analysis to identify datasets (RDDs) and transformations applied in the program. From this analysis, the module generates as output an intermediate representation of the program \textit{(c)} which is an implementation of the abstract model for data flow programs~\cite{sbmf2020}. 
In this model, a program is defined from a set of \textit{datasets}, a set of \textit{transformations} and a set of \textit{edges} that make the connection between the datasets and transformations, forming the data flow of the program.

The intermediate representation is passed as input to the \chl{MutationManager} module \textit{(d)}. This module is responsible for applying the mutation operators presented in Section~\ref{sec:mutation_operators} and generating a set of mutants \textit{(e)}. For a given mutation operator, the module checks whether it applies to the set of transformations. If so, it generates all the mutants that can be generated from this operator. 
Notice that each mutant has the same elements found in the original program, with modifications in one or more transformations.

The next steps of the process consist of generating executable versions of the mutants and executing them with the test set to analyze the results. As mentioned in Section~\ref{subsec:mutation_testing}, mutation testing can have a high computational cost due to the extensive amount of mutants that can be generated and executed in the process. Because of that, different works have proposed techniques to reduce the costs of mutation testing~\cite{Usaola2010}. In \transmut{} we apply two techniques for cost reduction: \textit{selective mutation}~\cite{Rothermel1993}, and \textit{Mutant Schema Generation} (MSG)~\cite{Untch1993}.

Selective mutation reduces the number of mutants that will be executed by removing redundant, equivalent, or trivial mutants (which are easily killed). In \transmut{}, this technique is applied by \chl{MutantReducer} \textit{(f)}. 
This module takes the set of mutants \textit{(e)} and applies \textit{reduction rules} to remove redundant mutants from the set.
These rules were stated in Section ~\ref{sec:mutation_operators} and they are summarized in Table~\ref{tab:reduction-rules}. 

The reduction rules UTDE, FTDS and OTDS are based on the fact that the operator UTD is a general case of the operators FTD, DTD and OTD  (UTDE). The operator UTD   deletes any unary transformation, whereas UTDE, FTDS and OTDS focus on specific transformation types. The rules also consider the relationship of the operator FTD with NFTP (FTDS) and the relationship of OTD with OTI (OTDS), as discussed in Section~\ref{sec:mutation_operators}. 

The reduction rule FTDS is related to the conditions in which mutants generated by the operators FTD and NFTP are killed. As discussed previously, even if these mutation operators generate different mutants for the same transformation, we experimentally observed that any test that kills a mutant generated by the operator FTD also kills a mutant generated by the operator NFTP operator applied in the same transformation. Thus, when FTD and NFTP are applied, the reduction rule FTDS removes mutants generated by the NFTP operator. The reduction rule OTDS follows the same principle for applying the operator  OTD that overrides the application of the operator OTI because its mutants will be killed whenever the mutants generated by OTD are killed. 

The reduction rules MTRR, DTIE and ATRC result from the analysis of the experimental results described in Section~\ref{sec:experiments}. These results empirically showed that some of the generated mutants do not lead to the design of new test cases. If the number of mutants had to be reduced, these mutants would be good candidates to be removed. The rule MTRR considers removing the mutants generated by the operator MTR that were trivial in most of our experimental cases. The rule DTIE addresses the removal of equivalent mutants generated by the operator DTI  that inserts the transformation \textit{distinct}  after aggregation and grouping transformations. 
Finally, the rule ATRC refers to the commutative replacement mutants generated by the ATR operator. This type of mutants is equivalent when the aggregation function is commutative. 

Note that the use of the module \chl{MutantReducer} is optional and configurable, so a user can select which reduction rules to apply according to the needs of his/her project.

\begin{table}[htbp!]
    \bigskip
	\centering
		\caption{Reduction rules applied in the module \chl{MutantReducer}.}
	\label{tab:reduction-rules}
\begin{tabular}{|c|p{.8\textwidth}|}
\hline
\textbf{Rule} & \textbf{Description}    \\ \hline
UTDE           & Removes mutants generated with the mutation operators FTD, DTD and OTD when the UTD operator has also been applied.      \\ \hline
FTDS           & Removes mutants generated with the mutation operator NFTP when the FTD or UTD operators have also been applied.    \\ \hline
OTDS           & Removes mutants generated with the mutation operator OTI when the OTD or UTD operators have also been applied. \\ \hline
MTRR           & Removes the following mutants generated with the MTR operator: mutants that map to $Max$ and $Min$, when the mapping is to a numerical type; mutants that map to `` '', when the mapping is to the type string; mutants that map to $x.reverse$, when the mapping is to a collection type; and mutants that map to $null$, when the mapping is to any other type. \\ \hline
DTIE           & Removes mutants generated with the mutation operator DTI when the \textit{distinct} transformation has been inserted after grouping or aggregation transformations. \\ \hline
ATRC           & Removes the commutative replacement mutants ($f_m(x,y) = f(y, x)$) generated with the ATR mutation operator.       \\ \hline         
\end{tabular}
\end{table}

The module \chl{MutantReducer} generates as output a new set of mutants \textit{(g)}, which is a subset of the original without redundant mutants.
On the next step, the set of mutants \textit{(g)} is passed as input to the module \chl{MetaMutantBuilder} \textit{(h)}. This module implements MSG, the second cost reduction technique applied in \transmut{}. In this technique, all the mutants generated for the program under test are incorporated into a single source program code, generating a ``\textit{meta-mutant}'' that incorporates all the mutants generated individually but that only needs to be compiled once. The MSG technique was used in \transmut{} because it is faster than other techniques~\cite{Offutt2001}, such as the interpretation-based technique that is used by classical mutation testing tools, such as \textit{Mothra}~\cite{DeMillo1989}, or the \textit{separate compilation} approach that is used by \textit{Proteum}~\cite{delamaro1996proteum}. Thus, the \chl{MetaMutantBuilder} module \textit{(h)} receives the mutant set \textit{(g)} as input, aggregates all mutants into a single program, and generates a meta-mutant as output \textit{(i)}.

Then, the meta-mutant and the class that implements the tests \textit{(j)} are passed to the module \chl{MutantRunner} \textit{(k)}. 
This module is responsible for managing the execution of the tests with the original program and mutants. 
If the original program tests fail, the tool ends the process and indicates that the original program or the tests need to be fixed. 
Otherwise, the tool executes the test set for each mutant and stores its results \textit{(l)}. 
These results are then passed to the \chl{MutantAnalyzer} module \textit{(m)}, which is responsible for analyzing the results of each mutant, checking whether the tests passed or failed, and returning the status (killed or lived) of each mutant \textit{(n)}. 
Finally, the results of the analysis are passed to the module \chl{Reporter} \textit{(o)} which is responsible for computing metrics, such as the number of mutants, the number of killed mutants, and mutation score, generating reports with the results of the process \textit{(p)}.

\subsection{Implementation details and architecture}\label{sec:detalhes}

The \transmut{} tool is implemented in \textit{Scala}~\cite{odersky2016}. We also chose Scala as the working language supported by the tool. Scala is a high-level language that incorporates object-oriented and functional programming; it is statically typed and executed in the JVM. 
We chose Scala for both tool development and supported language because it is the most used programming language for Spark, which offers better interaction with programs in that language since Spark was also developed in Scala~\cite{Zaharia:2010}.

\transmut{} was developed as a plugin for \textit{SBT} (\textit{Scala Build Tool})~\cite{Suereth2015}, a tool for building, managing, and deploying software projects in Scala and Java. 
SBT provides an interactive command-line interface that automates different software development tasks, including the compilation, tests, and publication of the project. 
SBT has the advantage of providing a simplified configuration language, increase productivity with automated build and test tasks, and can be easily extended through plugins~\cite{Suereth2015}. 
Developing the tool as a plugin allows direct access to the compiler and test execution components of SBT necessary to run the mutation testing process. 
Lastly, a plugin can be easily added to any project that uses SBT as a build tool, not requiring any additional installation, environment preparation, or specific operational system, making the tool portable and quickly adopted by different projects.

Figure~\ref{fig:transmut-architecture} presents an overview of the architecture of \transmut{}. The project is divided into six sub-projects that implement modules responsible for different mutation testing processes. Each sub-project is presented below:

\begin{figure}
	\centering
	\includegraphics[width=.7\textwidth]{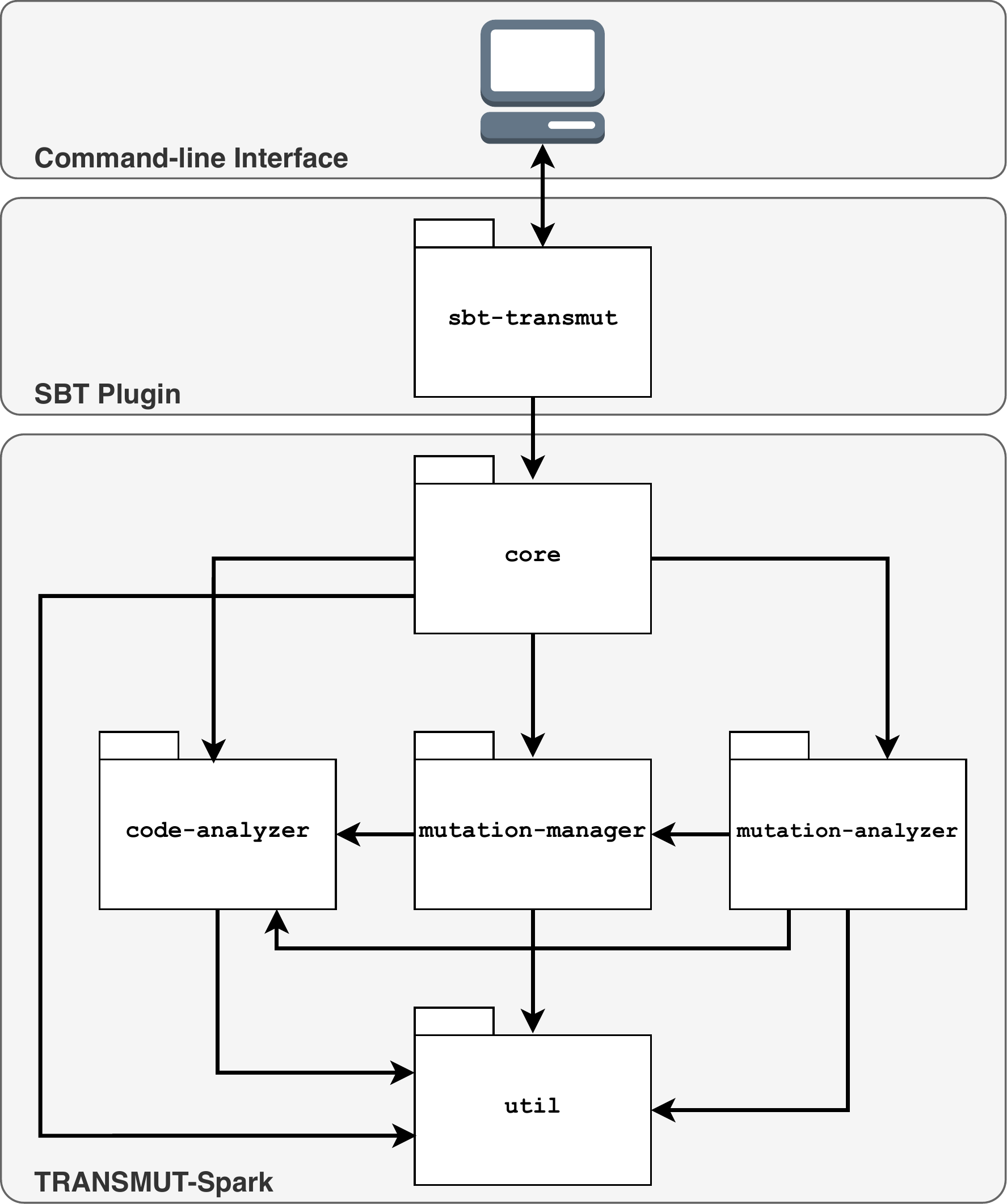}
	\caption{Overview of the \transmut{} architecture.}
	\label{fig:transmut-architecture}
\end{figure}

\paragraph*{\chl{util}:}
contains utility classes that are used by all other modules;
	
\paragraph*{\chl{code-analyzer}:} 
implements the module that analyzes the source code of a Spark program, doing a syntactic and type analysis, and generates an intermediate representation of this program as output;
	
\paragraph*{\chl{mutation-manager}:} 
contains the classes that implement the mutation operators presented in Section~\ref{sec:mutation_operators} and implements the modules for generating mutants, reducing mutants, implementing the reduction rules presented in Table~\ref{tab:reduction-rules}, and building the meta-mutant (that aggregates all the mutants generated into a single source code);

\paragraph*{\chl{mutation-analyzer}:} 
contains the interface to the module that manages the execution of the tests with the mutants and the original program, and the module that analyzes the results of the tests, indicating if the mutants were killed in the mutation testing process;
	
\paragraph*{\chl{core}:} 
main project that aggregates the other modules, implements the tool settings and defines the interface for the module that implements the mutation testing process. 
This module operates as a controller that defines the workflow of the other modules to execute the mutation testing process, being an implementation of the workflow presented in Figure~\ref{fig:transmut-process};
	
\paragraph*{\chl{sbt-transmut}:} 
implements the \transmut{} SBT plugin. 
This plugin defines the tasks that can be executed through the command-line interface of SBT. In addition, this project implements the concrete modules of the mutation testing process and test execution, dependent on the compiler and test components of SBT.
	
\bigskip 

For the development of the module \chl{ProgramBuilder}, that belongs to \chl{code-analyzer} and does the syntactic and type analysis of the program code to be tested, we used the \textit{Scalameta}\footnote{\url{https://scalameta.org}}, a library for reading, analyzing, and transforming Scala code. 
This library was also used to implement the mutation operators of \chl{mutation-manager} to make the changes in the source code.

For the development of the SBT plugin (\chl{sbt-transmut}), we take as reference the SBT plugin of the tool Stryker~\cite{stryker-ref}, an open-source tool for traditional mutation testing of programs in JavaScript, C\#, and Scala. 
We used as reference the test execution and plugin modules of \textit{Stryker} because these modules implement functionalities similar to the ones needed for our tool. 
Other modules of \transmut{} have been implemented differently from the modules of Stryker.

For the development of the module \chl{MetaMutantBuilder}, that belongs to \chl{mutation-manager} and receives a set of mutants as input and returns the meta-mutant as output, we used the technique of \textit{mutation switching} that is also applied by Stryker~\cite{stryker-ref}. 
This technique consists of putting all the mutants inside conditional expressions in the code and activating the mutant that must be executed through an environment variable. 
The component that controls the execution of the mutants is in charge of assigning values to that variable to indicate which mutant must be executed.

The project \transmut{}  has approximately 16k lines of code in Scala, including 10k lines of code of tests. 
\transmut{} is an open-source project distributed under the MIT License. 
Details on the use of the tool and its source code can be found in the repository: \url{https://github.com/jbsneto-ppgsc-ufrn/transmut-spark}.

\subsection{Use of the tool}\label{sec:utilizacao-transmut}
The following configuration steps must be done for executing \transmut. First, add it 
  as a plugin in a project that uses SBT and create a configuration file  \ct{transmut.conf} at the project's root. 
The file must contain at least the file names of the source code of the Spark program that will be transformed (\lstinline[style=Scala]|sources|), and the names of the methods that encapsulate the program (\lstinline[style=Scala]|programs|). 
The Spark program that will go through the mutation testing process with \transmut{} must be encapsulated in a method, so only that method is modified. 
Additional settings can be added to the file, such as the mutation operators (the tool applies all by default), test classes and reduction rules that will be applied.
Figure~\ref{fig:example-config} shows a configuration example for the program  in Figure~\ref{fig:wordcount-example}.

\begin{figure}
\centering
\begin{CenteredBox}
\begin{lstlisting}[style=Scala]
transmut {
	sources: [ "WordCount.scala" ],
	programs: [ "wordCount" ],
	test-only: [ "example.WordCountTest" ]
}
\end{lstlisting}     
\end{CenteredBox}
\caption{Example of a \transmut{} configuration file.}
\label{fig:example-config}
\end{figure}

\begin{figure}
	\centering
	\includegraphics[width=.95\textwidth]{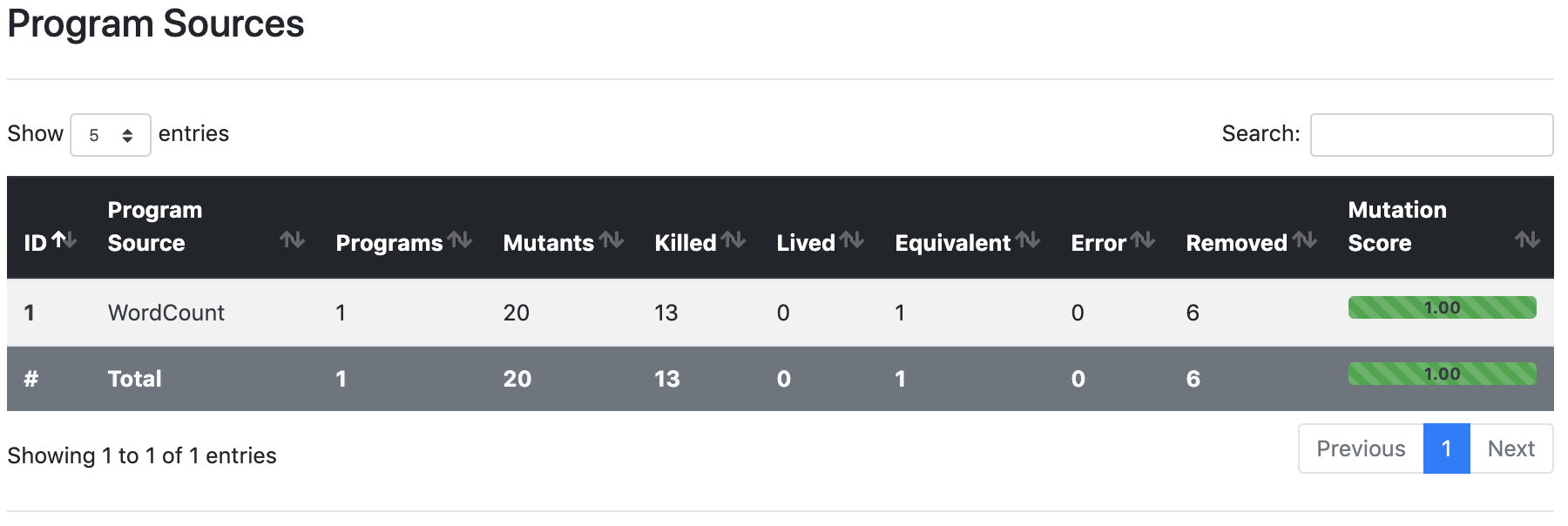}
	\caption{Part of the HTML report generated by \transmut{} with metrics about the programs.}
	\label{fig:transmut-report-1}
	\bigskip
\end{figure}

\begin{figure}
	\centering
	\includegraphics[width=.95\textwidth]{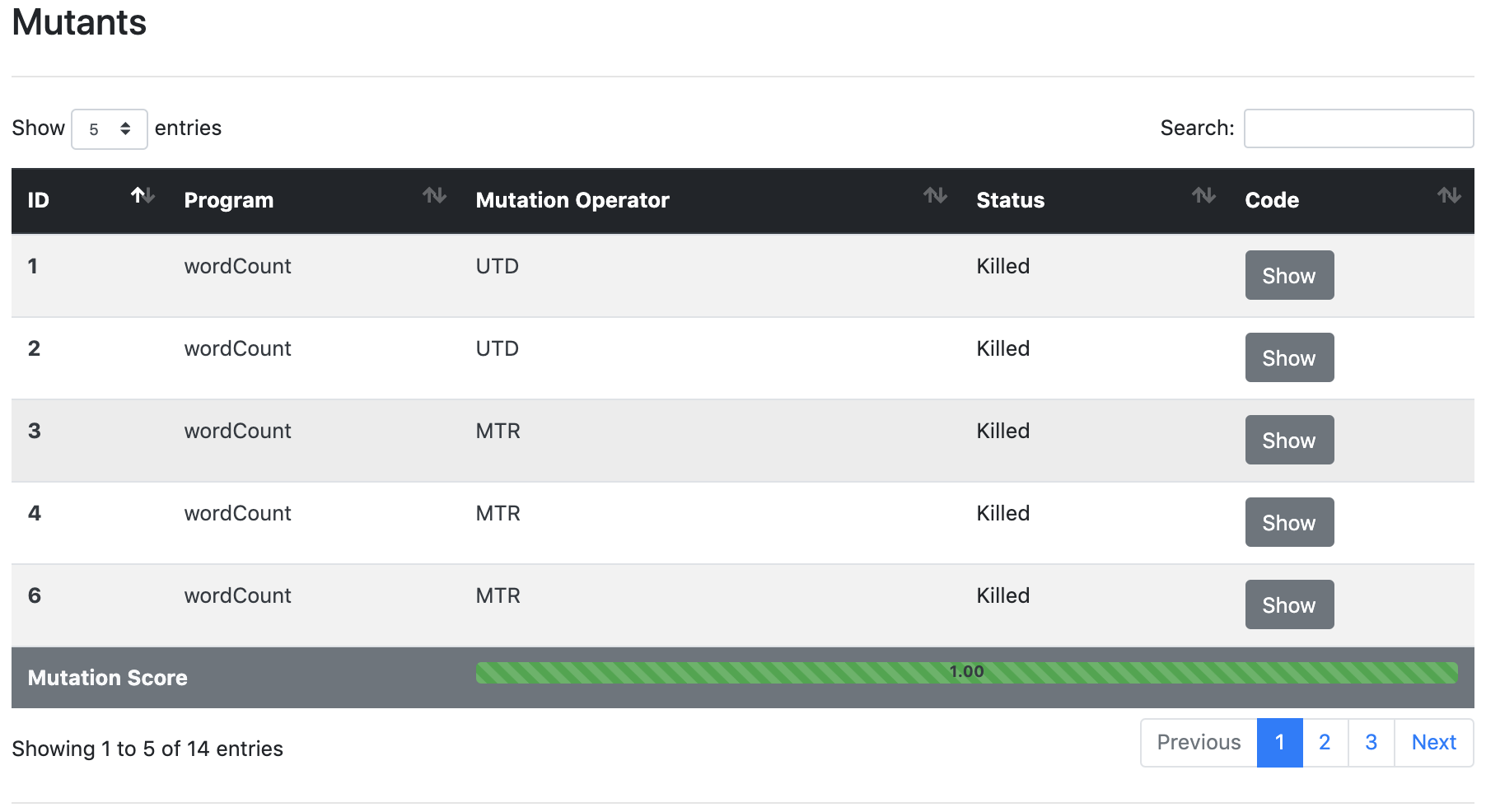}
	\caption{Part of the HTML report generated by \transmut{} with information about the generated mutants.}
	\label{fig:transmut-report-2}
\end{figure}

\begin{figure}
	\centering
	\includegraphics[width=.8\textwidth]{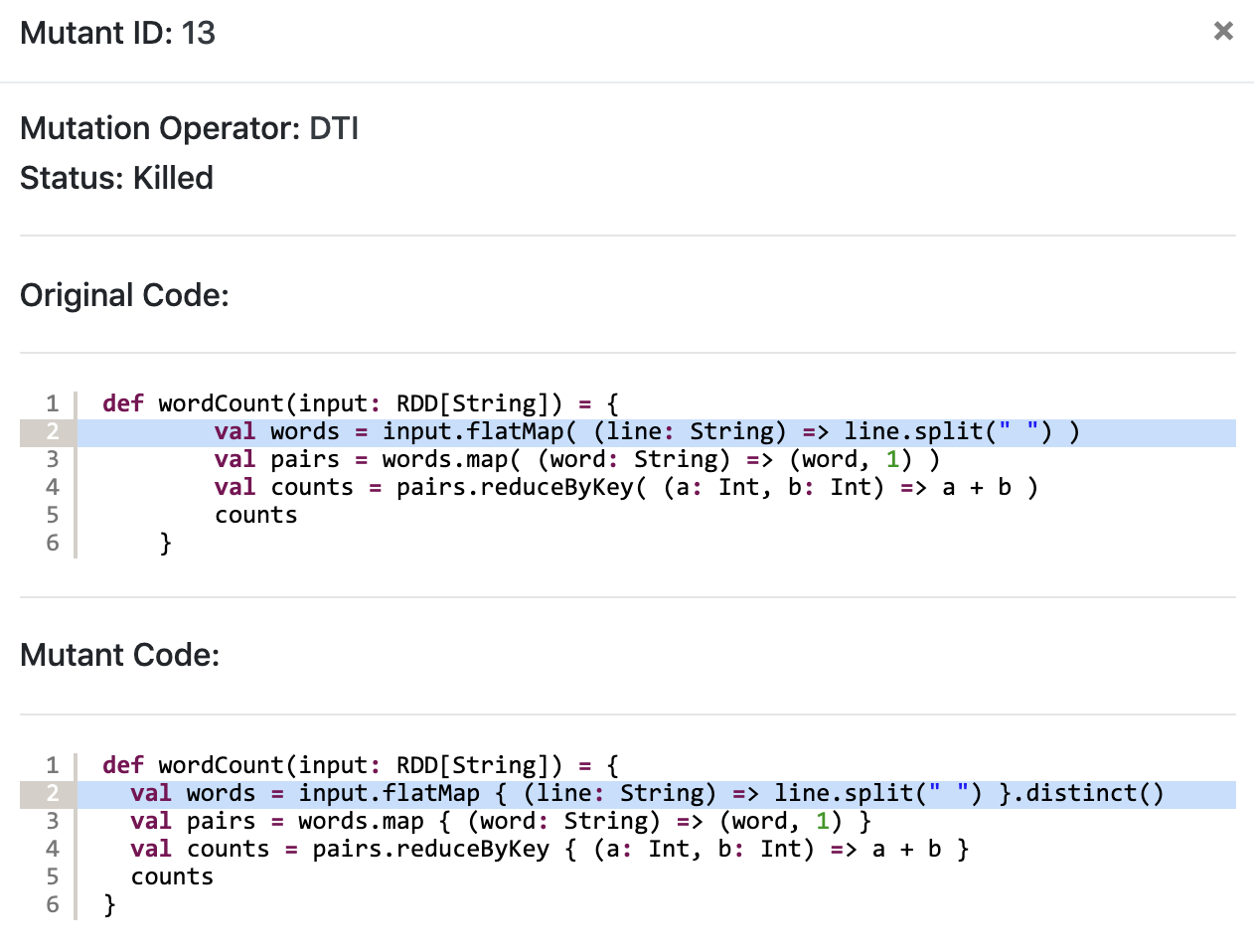}
	\caption{Part of the HTML report generated by \transmut{} with details of a mutant.}
	\label{fig:transmut-report-4}
\end{figure}

\begin{figure}
	\centering
	\includegraphics[width=.8\textwidth]{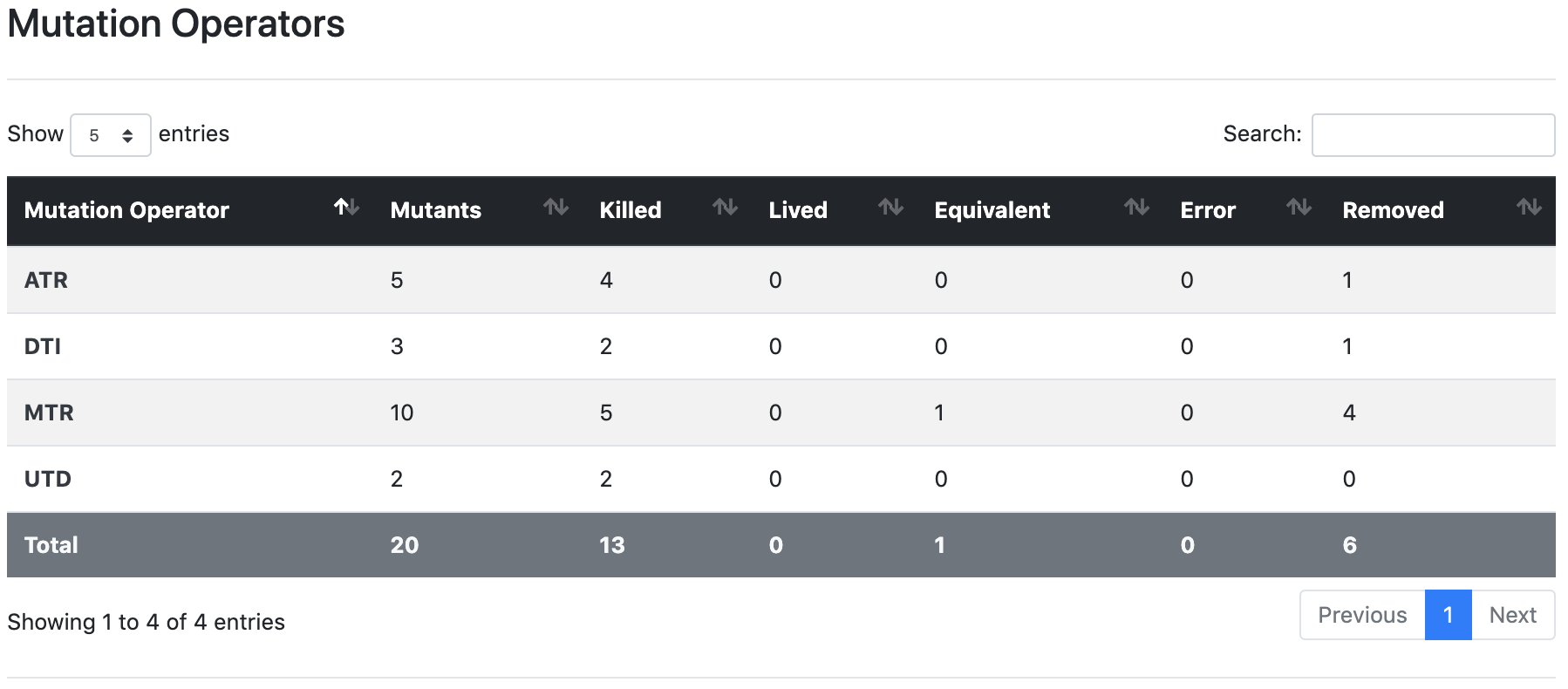}
	\caption{Part of the HTML report generated by \transmut{} with metrics from the mutation operators.}
	\label{fig:transmut-report-3}
\end{figure}

To execute \transmut{},  execute the command \ct{sbt transmut} from the project folder in the command-line terminal. 
This command triggers the execution of the mutation testing process following the workflow presented in Figure~\ref{fig:transmut-process}. 
When the execution terminates, \transmut{} generates reports with process results (HTML and JSON documents). 
The reports include the information necessary to complete the mutation testing process. 
Figures~\ref{fig:transmut-report-1},~\ref{fig:transmut-report-2},~\ref{fig:transmut-report-4} and~\ref{fig:transmut-report-3} present part of the HTML report generated by \transmut{} for the program presented in Figure~\ref{fig:wordcount-example}. 
 These report includes the metrics about the program (Figure~\ref{fig:transmut-report-1}); information about the generated mutants (Figure~\ref{fig:transmut-report-2}); details of a mutant, such as its status and the code of the original program and the mutant  (Figure~\ref{fig:transmut-report-4}); and general metrics about the mutation operators (Figure~\ref{fig:transmut-report-3}).

The reports show the mutants that are alive, and they can be analyzed to verify if they are equivalent. 
Identifying equivalent mutants allows verifying if the mutation testing process must continue to achieve the established mutation score. 
To support the whole process, \transmut{} has the command \ct{transmutAlive} that causes the process to be executed again only for the mutants that are alive of the last execution—this command tags in the report the mutants that have been identified as equivalent. 
The identification of equivalent mutants is made by inserting their identifiers in the tool settings in the field\lstinline[style=Scala]|equivalent-mutants|. This command forces the execution of mutants identified as relevant but removed by the reduction module.
It also allows new tests to be inserted into the test class to kill non-equivalent mutants that are still alive. When using this command, a new report is generated with the updated metrics and results. More details on the use of the tool and its settings can be found in its repository.

\section{Experimental Evaluation}\label{sec:experiments}
This section presents the experiments we conducted to evaluate \transmut. These experiments are a variation of the experiments presented in a previous paper~\cite{caise2020} which were intended to validate the transformation mutation approach, not the tool. 
  Therefore, the mutation testing process steps, including the generation and execution of mutants, were performed manually. In contrast, the objective of the experiments described in this section is to evaluate the  \transmut{}  that automates the mutation testing process.
  The experimental validation was designed 
(i) to evaluate the tools' performance in terms of its capacity to automate the most laborious steps of the mutation testing process (in contrast to the manual experiments~\cite{caise2020});
  (ii) compare its performance when reduction rules were applied, and the tests were performed with the ``minimum'' amount of mutants; 
(iii) compare the performance and determine to which extent of \transmut{} an existing Scala programs mutation tools are complementary (see Section \ref{sec:transmut-vs-scalamu}).

\subsection{Goals and Methodology}
The goals of the experiments are (i) to analyze the applicability and evaluate the costs of using \transmut{} in Spark programs' mutation testing process; (ii)  evaluate the impact of the mutants reduction module; (iii) the pertinence of mutation operators used in the mutation process concerning the quality of the resulting mutation testing. 
To achieve these goals, we used  \transmut{} to apply mutation testing to a set of nine representative Spark programs as described in Figure~\ref{fig:mutation_testing_process}. 
For eight of the programs in the testing battery, we used the tests developed for the manual experiments~\cite{caise2020}. The tests were designed to kill non-equivalent mutants, searching to achieve a mutation score (\textit{ms}) of $1.0$.

We added the program \texttt{MovieLensExploration} to the testing battery with a  complex architecture that makes it unviable to perform manual mutation testing due to the number of generated mutants.  Indeed, the mutation process generates for the program \texttt{MovieLensExploration} as many mutants as the total number of mutants generated for the eight programs of the testing battery together. Similar to the other programs of the testing battery, we developed tests searching to achieve an \textit{ms} of $1.0$.

Figure~\ref{fig:mutation_testing_process} shows the adopted experiment methodology, with  \transmut{} automating the generation, execution and analysis of mutants.  The design of tests to kill the mutants and conclusions from provided reports are still a manual activity. 

\paragraph{\em Dealing with complex program structures (programs and subprograms)}
Automating  mutation tasks thanks to 
\transmut{} let adopt a specific mutation generation strategy for testing programs composed of subprograms (methods). In our previous manual experiment~\cite{caise2020},  transformations for generating mutants considered subprograms as a single program. Thus mutation operations were applied considering the set of subprograms' as a whole.
 Besides, this strategy let transformations in two different subprograms be exchanged or replaced using the mutation operators.
 \transmut{}, in contrast,  treats each subprogram as an independent program, with its own set of transformations. Thus,   mutation operators also considered the set of transformations of each subprogram separately, enabling changes and substitutions of transformations among the transformations in the subprogram. This choice had an impact on the number of generated mutants for each program.

\paragraph{\em Experimental process}
 was designed to assess the reduction module and compare results with the manual experiments~\cite{caise2020}, as said above. Therefore,  the mutation process with \transmut{} was executed twice, alternatively enabling and disabling the reduction module.

\paragraph{\em Assessment metrics}
We adopted the following metrics for comparing  the three  experiments:
\begin{enumerate}
    \item the time spent by the test developer in the process; 
    \item  the mutants execution and analysis;
    \item the occurrence of errors in the process;
    \item the execution time of the tool for testing experiment cases; 
    \item the  \textit{Killed Ratio} (\textit{KR}) to assess each  mutation operator. 
\end{enumerate}

 The  \textit{KR}  gives insights about the extent to which mutation operators can generate mutants difficult to kill, \textit{i.e.}, that simulate faults that would be more difficult to detect. 
For a set of tests developed to achieve the mutation score of 1.0, \textit{KR} corresponds to the ratio between the number of tests that killed the generated mutants and the total number of tests executed with those mutants. Low KR values show that mutants generated by a specific operator were killed by fewer tests, meaning that they were hard to kill. In contrast, high KR values indicate that mutants were easily killed.

\paragraph{Testing battery}
used in experiments consists of programs implementing representative types of Big Data processing tasks such as analysis of texts and log files, queries in tabular databases based on the queries presented in~\cite{AMPLab}, and data recommendation using the \textit{collaborative filtering} algorithm~\cite{Sarwar:2001}. The following lines briefly describe the programs applied in the evaluation\footnote{The programs used in the experiments of this work are publicly available at \url{https://github.com/jbsneto-ppgsc-ufrn/spark-mutation-testing-experiments}.}.

\begin{itemize}
    \item {\em Exploring tabular datasets:} The programs \texttt{ScanQuery}, \texttt{AggregationQuery}, \texttt{JoinQuery} and \texttt{DistinctUserVisitsPerPage} explore tabular datasets containing websites data such as their ranking and visiting users. They implement the queries introduced in the \textit{AMPLab Big Data Benchmark}~\cite{AMPLab}, which is a benchmark for large data processing systems. 
    
    \item {\em Analysing textual datasets:} The programs \texttt{NGramsCount} and \texttt{NasaApacheWebLogsAnalysis} analyse text datasets computing the frequency of n-grams and analysing log messages to identify unique messages in different files.
    
    \item {\em Analysing data thorugh programs with ``complex'' architectures (programs and subprograms):} The programs \texttt{MoviesRatingsAverage}, \texttt{MoviesRecomendation} and \texttt{MovieLensExploration}  analyse  datasets produced by the application \textit{MovieLens}~\cite{Harper:2015}. It is  a system where users can rate movies and receive recommendations based on their preferences. The program \texttt{MovieLensExploration} composed of eight subprograms, performs a series of exploratory analyzes on the datasets of \textit{MovieLens}.

\end{itemize}

\subsection{Results}
These experiments were performed on a MacBook Pro with a 1,4 GHz Intel Core i5 processor, 16 GB of RAM (2133 MHz LPDDR3) and a 256 GB SSD.
Table~\ref{tab:results_programs_experiments} aggregates the experiment results for each tested program and groups them concerning the manual \cite{caise2020} (the column ``First Experiment'') and automatic mutation testing results with the reduction module disabled (``Second Experiment'') and enabled (``Third Experiment'').
For each program, the table shows the number of Spark transformations applied in the program (\textit{Transf}.), the number of tests developed for the program (\textit{Tests}), the number of generated mutants  (\textit{\#M}), the number of killed mutants  (\textit{\#K}) and the number of equivalent mutants (\textit{\#E}). Additionally, for the results obtained with the tool, the table shows the tool's execution time in seconds (\textit{Time (s)}) and the number of mutants removed by the reduction module (\textit{\#R}). 
The tool's total execution time is given by the time spent generating mutants, executing tests and generating reports. 

\begin{table}[!htbp]
\caption{Results of the experiments aggregated by program.}
\label{tab:results_programs_experiments}
\begin{adjustbox}{width=1\textwidth}
\begin{tabular}{|c|c|c|c|c|c|c|c|c|c|c|c|c|c|c|}
\hline
\multicolumn{3}{|c|}{\textbf{}}                               & \multicolumn{3}{c|}{\textbf{First Experiment}} & \multicolumn{4}{c|}{\textbf{Second Experiment}}                & \multicolumn{5}{c|}{\textbf{Third Experiment}}              \\ \hline
\textbf{Program}          & \textbf{Transf.} & \textbf{Tests} & \textbf{\#M}   & \textbf{\#K}  & \textbf{\#E}  & \textbf{\#M} & \textbf{\#K} & \textbf{\#E} & \textbf{Time (s)} & \textbf{\#M} & \textbf{\#K} & \textbf{\#E} & \textbf{\#R} & \textbf{Time (s)} \\ \hline
\texttt{NGramsCount}               & 5                & 5              & 27             & 22            & 5             & 32           & 27           & 5            & 243.2             & 23           & 22           & 1            & 9            & 176.8             \\ \hline
\texttt{ScanQuery}                 & 3                & 3              & 12             & 12            & 0             & 13           & 13           & 0            & 97.6              & 7            & 7            & 0            & 6            & 57.6              \\ \hline
\texttt{AggregationQuery}          & 3                & 3              & 15             & 13            & 2             & 16           & 14           & 2            & 132.4             & 10           & 10           & 0            & 6            & 90.8              \\ \hline
\texttt{DistinctUserVisitsPerPage} & 4                & 2              & 16             & 10            & 6             & 17           & 11           & 6            & 142.6             & 11           & 7            & 4            & 6            & 98.2              \\ \hline
\texttt{MoviesRatingsAverage}      & 5                & 4              & 25             & 22            & 3             & 19           & 14           & 5            & 182.8             & 13           & 11           & 2            & 6            & 135.6             \\ \hline
\texttt{MoviesRecomendation}       & 12               & 5              & 37             & 33            & 4             & 56           & 41           & 15           & 524.4             & 35           & 24           & 11           & 21           & 355.2             \\ \hline
\texttt{JoinQuery}                 & 11               & 6              & 27             & 25            & 2             & 37           & 35           & 2            & 332.6             & 21           & 20           & 1            & 16           & 201.8             \\ \hline
\texttt{NasaApacheWebLogsAnalysis} & 7                & 4              & 55             & 49            & 6             & 38           & 28           & 10           & 323.6             & 31           & 21           & 10           & 7            & 286.2             \\ \hline
\textbf{Total}            & \textbf{50}      & \textbf{32}    & \textbf{214}   & \textbf{186}  & \textbf{28}   & \textbf{228} & \textbf{183} & \textbf{45}  & \textbf{1979.2}   & \textbf{151} & \textbf{122} & \textbf{29}  & \textbf{77}  & \textbf{1402.2}   \\ \hline
\end{tabular}
\end{adjustbox}
\end{table}

Table~\ref{tab:results_operators_experiments} presents the results of the eight programs aggregated by the mutation operator. In it, we can see the number of generated mutants (\textit{\#M}), the number of equivalent mutants (\textit{\#E}) and the KR metric (\textit{KR (\%)}), as well as the number of mutants removed for the experiments with the reduction module enabled in the tool (\textit{\#R}).

\begin{table}[!htbp]

\caption{Results of the experiments aggregated by mutation operator.}
\label{tab:results_operators_experiments}
\begin{CenteredBox}
\begin{tabular}{|c|c|c|c|c|c|c|c|c|c|c|}
\hline
\textbf{}         & \multicolumn{3}{c|}{\textbf{First Experiment}} & \multicolumn{3}{c|}{\textbf{Second Experiment}} & \multicolumn{4}{c|}{\textbf{Third Experiment}} \\ \hline
\textbf{Operator} & \textbf{\#M} & \textbf{\#E} & \textbf{KR (\%)} & \textbf{\#M}  & \textbf{\#E} & \textbf{KR (\%)} & \textbf{\#M}  & \textbf{\#E}  & \textbf{\#R}  & \textbf{KR (\%)} \\ \hline
UTS               & 11           & 2            & 67.6             & 4             & 2            & 25.0             & 4             & 2             & 0             & 25.0             \\ \hline
BTS               & 1            & 0            & 75.0             & 0             & 0            & --               & 0             & 0             & 0             & --               \\ \hline
UTR               & 22           & 2            & 39.0             & 8             & 2            & 37.5             & 8             & 2             & 0             & 37.5             \\ \hline
BTR               & 2            & 0            & 37.5             & 0             & 0            & --               & 0             & 0             & 0             & --               \\ \hline
UTD               & 6            & 0            & 32.0             & 19            & 0            & 32.9             & 19            & 0             & 0             & 32.9             \\ \hline
MTR               & 82           & 5            & 76.1             & 91            & 9            & 73.8             & 48            & 4             & 43            & 67.1             \\ \hline
FTD               & 7            & 0            & 34.4             & 7             & 0            & 36.1             & 0             & 0             & 7             & --               \\ \hline
NFTP              & 7            & 0            & 65.6             & 7             & 0            & 66.6             & 0             & 0             & 7             & --               \\ \hline
STR               & 10           & 2            & 34.4             & 8             & 0            & 34.4             & 8             & 0             & 0             & 34.4             \\ \hline
DTI               & 31           & 10           & 27.7             & 49            & 25           & 26.2             & 42            & 18            & 7             & 26.2             \\ \hline
DTD               & 1            & 0            & 25.0             & 1             & 0            & 25.0             & 0             & 0             & 1             & --               \\ \hline
ATR               & 20           & 4            & 46.4             & 20            & 4            & 46.4             & 16            & 0             & 4             & 46.4             \\ \hline
JTR               & 6            & 3            & 22.2             & 6             & 3            & 22.2             & 6             & 3             & 0             & 22.2             \\ \hline
OTI               & 4            & 0            & 30.0             & 4             & 0            & 25.0             & 0             & 0             & 4             & --               \\ \hline
OTD               & 4            & 0            & 20.0             & 4             & 0            & 16.6             & 0             & 0             & 4             & --               \\ \hline
\end{tabular}
\end{CenteredBox}
\end{table}

Table~\ref{tab:results_operators_movielens} shows the aggregated results by mutation operator of the mutants generated for the program \texttt{MovieLensExploration}. \transmut{} applied
33 Spark transformations for the program \texttt{MovieLensExploration}    composed of eight subprograms with the reduction module disabled. 
In the second evaluation, with the reduction module enabled, \transmut{} generated 195 mutants with 24 identified as equivalent. We  developed 17 test cases to reach an \textit{ms} of $1.0$. The average execution time of the tool for this program was approximately 39.4 minutes (2364.6 seconds). In the third experiment, with \transmut{} using the mutants reduction module, the tool removed 65 of the mutants generated in the second evaluation, leaving a total of 130 mutants out of which 11 were equivalent. With this reduction, the tool's average execution time was approximately 24.6 minutes (1475.8 seconds).

\begin{table}
\caption{Results of the experiments for the program \texttt{MovieLensExploration} aggregated by mutation operator.}
\label{tab:results_operators_movielens}
\begin{CenteredBox}
\begin{tabular}{|c|c|c|c|c|c|c|c|}
\hline
\textbf{}         & \multicolumn{3}{c|}{\textbf{Second Experiment}} & \multicolumn{4}{c|}{\textbf{Third Experiment}}                \\ \hline
\textbf{Operator} & \textbf{\#M}  & \textbf{\#E} & \textbf{KR (\%)} & \textbf{\#M} & \textbf{\#E} & \textbf{\#R} & \textbf{KR (\%)} \\ \hline
UTS               & 3             & 0            & 88.8             & 3            & 0            & 0            & 88.8             \\ \hline
UTR               & 6             & 0            & 94.4             & 6            & 0            & 0            & 94.4             \\ \hline
UTD               & 8             & 0            & 69.2             & 8            & 0            & 0            & 69.2             \\ \hline
MTR               & 114           & 10           & 99.2             & 66           & 5            & 48           & 98.6             \\ \hline
DTI               & 32            & 11           & 43.4             & 25           & 6            & 7            & 42.8             \\ \hline
ATR               & 20            & 3            & 73.9             & 16           & 0            & 4            & 72.7             \\ \hline
JTR               & 6             & 0            & 22.2             & 6            & 0            & 0            & 22.2             \\ \hline
OTI               & 3             & 0            & 58.3             & 0            & 0            & 3            & --               \\ \hline
OTD               & 3             & 0            & 41.6             & 0            & 0            & 3            & --               \\ \hline
\end{tabular}
\end{CenteredBox}
\end{table}

\subsection{Analysis and Discussion}
For the analysis of the results, we compare the results of the first experiment with the results obtained using the \transmut{} without the reduction module (second experiment). Then, we compare the results obtained with \transmut{} with and without using the reduction module (third experiment).

\begin{figure}
	\centering
	\includegraphics[width=\textwidth]{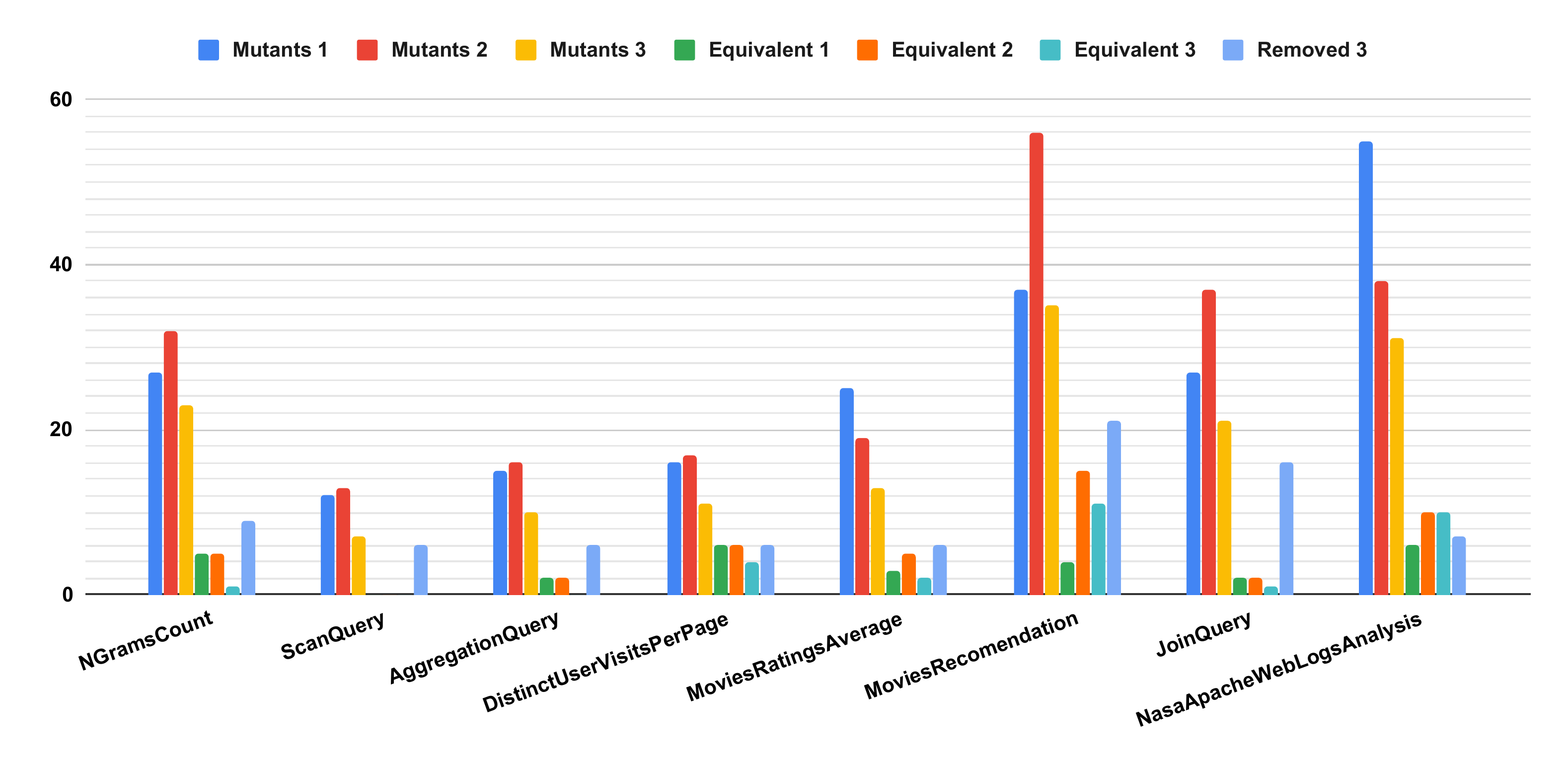}
	\caption{Comparison of aggregated results by program.}
	\label{fig:chart-differences-programs}
\end{figure}

\begin{figure}
	\centering
	\includegraphics[width=\textwidth]{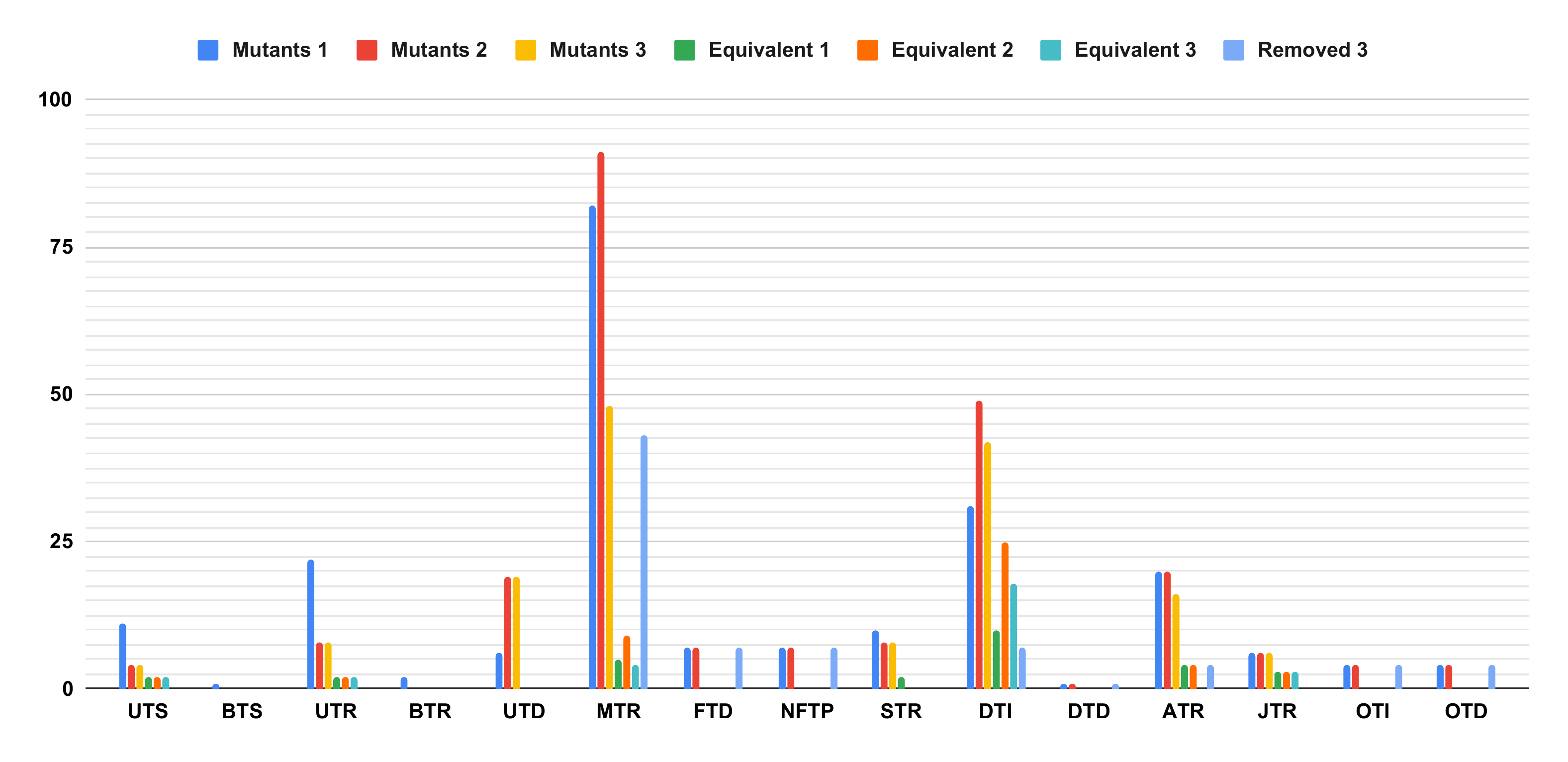}
	\caption{Comparison of aggregated results by mutation operator.}
	\label{fig:chart-differences-operators}
\end{figure}

Figure~\ref{fig:chart-differences-programs} shows the results of the three experiments aggregated by  program. The figure compares the total number of mutants and equivalent mutants generated in the first (see columns \textit{Mutants 1} and \textit{Equivalent 1}) and second experiments (see columns \textit{Mutants 2} and \textit{Equivalent 2}). Then columns \textit{Mutants 3} and \textit{Equivalent 3} compare the number of mutants and equivalent mutants generated with \transmut{}  using the reduction module. Finally, column \textit{Removed 3} shows the number of removed mutants by the reduction module.
The differences between the results of the three experiments aggregated by mutation operator can be seen in Figure~\ref{fig:chart-differences-operators}.

\paragraph{\em Analysis of the number of mutants}
Figure~\ref{fig:chart-differences-programs} shows that in most cases, the second experiment generated more mutants than the first experiment: 14 more mutants in general and 17 additional equivalent mutants. 
The exceptional case regards the programs \texttt{MoviesRatingsAverage} and \texttt{NasApacheWebLogsAnalysis}.

For five programs the number of equivalent mutants remained the same: \texttt{NGramsCount}, \texttt{ScanQuery}, \texttt{AggregationQuery}, \texttt{DistinctUserVisitsPerPage} and \texttt{JoinQuery}. The number of equivalent mutants was bigger in the second experiment for  three programs: \texttt{MoviesRatingsAverage}, \texttt{MoviesRecomendation} and \texttt{NasaApacheWebLogsAnalysis}.

Comparing the results of the first and second experiments regarding the programs \texttt{MoviesRecomendation} and \texttt{JoinQuery}, the second experiment generated around 
51\% and 37\%, respectively, more mutants than in the first one. 
These results show that with  \transmut{} we could generate mutants that were not generated in the manual process.
Regarding the programs \texttt{NasApacheWebLogsAnalysis} and \texttt{MoviesRatingsAverage}, the second experiment generated around 30\% and 24\% fewer mutants than the first experiment.
This reduction in the number of mutants is due to technical strategies implemented by \transmut{}  regarding how it transforms programs composed of subprograms (methods).  Recall that \transmut{} treats subprograms as independent programs for generating mutants. In consequence, as shown in Figure~\ref{fig:chart-differences-operators} the operators UTS, BTS, UTR and BTR generate fewer mutants.

In the second experiment, the data flow mutation operators (UTS, BTS, UTR, BTR and UTD) generated approximately 13.6\% of the mutants for the eight programs of the testing battery. In contrast, in the first experiment, they generated  20\% of the mutants. 
The reduction of mutants for the operators UTS, BTS, UTR, and BTR had no negative impact on the results of the second experiment because the tests that killed their mutants were also responsible for killing mutants of other operators.

Besides the differences between the programs and subprograms handling in the tool, other differences between results in both experiments can be due to human errors in the manual generation of mutants in the first experiment.
As shown in  Figure~\ref{fig:chart-differences-operators}, human errors impact the results of the operators UTD, MTR and DTI.
Indeed, the mutant generation process calls for a detailed analysis of target programs to identify possible modification points within the code the lines where mutation operators can be applied. This process is challenging given the number of generated mutants, making a manual generation prone to errors. 

Thus, we observed that the differences concerning the operators UTD, MTR and DTI occurred due to errors in the manual mutant generation process. These errors explain a smaller number of mutants created for these operators in the first experiment. Concerning the other mutation operators, the number of mutants generated was consistent in both experiments (with a slight difference for the STR operator where we eliminated equivalent mutants while implementing the tool). 

Regarding the program \texttt{MovieLensExploration},  a total of 195 mutants were generated, 24 of which were equivalent.  
This amount corresponds to approximately 85\% of the total number of mutants generated for the other eight programs used in the experiment. 

\paragraph{\em Analysis of the metric KR}

 Mutation operators did not have significant variations between the first and second experiments regarding KR. Two exceptions observed are 
(1) the operators BTS and BTR that did not generate mutants with \transmut{} due to the difference in the manipulation of programs and subprograms in the first experiment (manual) and the tool; and 
(2) UTS and UTR generated fewer mutants and had more significant variations in their KR's.
 
The operators with better results were UTS, DTI, DTD, JTR, OTI and OTD with low KR values (below 30\%). These values show that a smaller amount of tests killed the mutants generated by them.
In contrast, the mutants generated by operators with high KR values, like MTR and NFTP, had their mutants killed by a more significant amount of tests. This intuitively indicates that these operators might be trivial.

As shown in Table~\ref{tab:results_operators_movielens}  the KR values of the program \texttt{MovieLensExploration} imply that the mutants generated with the operators MTR, UTS and UTR were killed by most tests developed for this program.
As in the first experiment, the mutants generated with the operator MTR  were the ones that died more easily and contributed less in the test generation process since they did not need particular tests to kill them. In contrast, the operators DTI, JTR and OTD  had the lowest KR, indicating that their mutants required more specific tests to be killed.

\paragraph{\em Analysis of the generated mutants per operator}
We observed that the mutants generated by MTR were trivially killed in most cases where the mutants were mapped to $Max$, $Min$, `` '', $x.reverse$ and $null$. In contrast, MTR mutants 
mapped to other values had better results. 

As discussed in  Section~\ref{sec:mutation_operators} through the experiments, we confirmed the relation between couples of operators FTD/NFTP and OTD/OTI that shows that the tests that kill FTD/OTD mutants always kill NFTP/OTI mutants, but not the opposite.
Finally, since the operator UTD  represents a general transformation removal, the mutants generated with the FTD, DTD and OTD operators were also generated by UTD. Thus, the application of the operator UTD overrides the application of FTD, DTD and OTD; otherwise, mutants are duplicated. These results were used to define the reduction rules (see Table~\ref{tab:reduction-rules}) implemented by the mutants reduction module.

Table~\ref{tab:results_operators_movielens}  shows the number of mutants generated for each mutation operator applied to the program \texttt{MovieLensExploration}. Note that the operator MTR generated approximately 58\% of the program's mutants. Of the 33 transformations applied in the program \texttt{MovieLensExploration}, 20 were mapping transformations (approximately 60\%).

The more significant amount of mapping transformations explains the difference between the operator MTR  and the other operators, where the number of mutants is proportional to the type of transformations. As in the first experiment, the operator DTI  generated the second largest number of mutants and the most significant number of equivalent mutants. The other mutation operators generated mutants proportional to the number of transformations performed on each target type within the program.

\paragraph{\em Analysis of the execution time of the mutation process}
The most significant difference between the first two experiments was the time spent executing the mutation testing process. The experiments' execution and analysis of the first experiment results took approximately four weeks, with an average of approximately three days for each program. As previously mentioned, this work involved the manual generation of the mutants, implementation of test cases, implementation of scripts for automatic execution of the mutants, and manual collection and analysis of the results, being most of these laborious and repetitive tasks.

The second experiment tested \transmut{} and showed that the time spent in the execution of the mutation testing process was drastically reduced. The tool automates the tasks of generating the mutants, executing the tests and analyzing the results. Thus, efforts were concentrated in developing test cases and the analysis of living mutants to identify equivalent ones. In the end, the work that took on average three days for each program in the first experiment was done in a few hours in the second experiment. This effort was significant for the program
\texttt{MovieLensExploration} that was not applied in the first experiment due to its complexity compared to the other programs, being impractical to apply the process manually. Thus, with the aid of \transmut{}, the effort in the mutation testing process of \texttt{MovieLensExploration} was concentrated on the development of tests, which remains manual.

For the eight programs in the first experiment, the process with \transmut{}, without taking into account the time spent with the development of the tests since they were reused from the first experiment, took a few minutes for each program (see Table~\ref{tab:results_programs_experiments}). The program \texttt{MoviesRecomendation} was the one that generated the most significant amount of mutants and took the longest time to finalize the process (approximately nine minutes). Regarding the program \texttt{MovieLensExploration}, we spent approximately one day developing its tests, and the execution of the process took approximately 39 minutes on average.

\subsection{Lessons learned and Discussion}
\noindent
 {\em Lessons learned.}\\
The analysis of the results of the first and second experiments motivated the development of the reduction rules (MTRR, DTIE, ATRC, FTDS, OTDS and UTDE) introduced in Section \ref{sec:mutation_operators} and the implementation of the reduction module of \transmut{}. This module applies a \textit{selective mutation} strategy~\cite{Rothermel1993} to remove equivalent, redundant or trivial (i.e., mutants that are easily killed) mutants from the set of mutants generated by \transmut{}. This strategy reduces the number of mutants to be executed and not contribute to the mutation testing process. 

Table~\ref{tab:results_programs_experiments} and table~\ref{tab:results_operators_experiments}
show 
the results obtained by  \transmut{} using the reduction module on the eight programs of the testing battery (third experiment). Recall that the same testing battery was used in the first and second experiments. 
 Figure~\ref{fig:chart-differences-programs} and Figure~\ref{fig:chart-differences-operators}
 show the experimental result with the reduction model disabled and enabled.
Note that the reduction module removed approximately 34\% of the mutants generated by \transmut{} when the module was disabled. The module removed approximately 35\% of equivalent mutants. The reduction of equivalent mutants reduces the effort required to analyze living mutants since equivalent mutants need to be detected manually.

The programs \texttt{MoviesRecomendation} and \texttt{JoinQuery}  had  approximately  37\% and 43\% fewer mutants than the second experiment, respectively.  The mutants generated with the operators FTD, DTD, OTD, NFTP and OTI  were removed using the reduction rules UTDE, FTDS and OTDS. Then, approximately 47\% of the mutants generated by the operator MTR were removed. For the operators DTI and ATR, all mutants removed by the module were equivalent.

The KR metric improved for the mutation operator MTR, reducing about 6.7\%. 
The KR for the operators DTI and ATR  did not change, only equivalent mutants were removed, and the metric is not affected in this case. Finally, the removed mutants caused by the operators FTD, DTD, OTD, NFTP and OTI had no adverse effects on the process because their mutants were redundant with the mutants generated by the operator UTD. The application of these redundant mutants was unnecessary because the operator UTD was applied.

For the program \texttt{MovieLensExploration}, the reduction module removed approximately 33\% of the mutants generated by \transmut{}. In the case where only equivalent mutants are considered, this reduction was around 46\%. 
As for the other eight programs, the most significant reduction concerned the operator MTR, with 46\% fewer mutants. The KR of the operators MTR, DTI and ATR dropped approximately 1\% compared to the KR obtained without the reduction module. Even if this variation was slight, the result is promising because it shows that the module could remove inefficient mutants from the process without harmful side effects like not detecting failures.

One of the major impacts of the mutants reduction module was on the execution time. For the eight programs used in the first experiment, the module reduced the total execution time of the tool by 29\%. For the program \texttt{MovieLensExploration}, this reduction was 37\%. These reductions in the tool's execution time were proportional to the number of mutants removed. In general, the third experiment's results show that the mutants reduction module improved \transmut{} results.  The module could reduce the number of equivalent mutants and thereby the effort required for identifying them. The module could remove redundant and trivial mutants.  
The module improved the KR  results for some mutation operators, meaning that it removed inefficient mutants. Finally, the module reduced the tool's execution time. Thus, the mutants reduction module achieved its goal by reducing the costs of the mutation testing process of  \transmut{}.

\noindent
 {\em Discussion.}\\
The main goal of the experiments was assessing the use of a testing tool for testing Spark programs (when comparing with a less systematic and error-prone manual process). We also verified to which extent the tool allows more extensive, more realistic tests.

The manual experiment let us define a baseline that serves as a reference to assess two main aspects regarding the tool: 
\begin{enumerate}
    \item [1] Automation makes the mutation testing process more agile. It prevents programmers or people testing the tool to perform manual tasks that are far away from the objective of the task, which is assessing Spark data processing code. Once with the baseline, it is possible to determine to which extent an agile and automated process can lead to the generation of more mutants, calibrate the reduction strategy, easily compare the testing results, eventually activate/deactivate mutants and thoroughly understand and calibrate the testing process. Our results show that it is essential to have a tool with the facilities offered by \transmut\ within a testing process. 
    
    \item  [2] Keeping quantitative track of the testing process.\transmut\ also produces statistics that are useful to generate quantitative results about the testing process with metrics that can be representative to compare testing tasks under different conditions. A manual process would imply that programmers drag the pencil with a considerable burden to produce these testing results.
\end{enumerate}

Finally, concerning mutation operators through the tool's implementation, we could profile the set of operators, identifying those that override specific ones. This result let us define a ``minimum'' set of mutation operators adapted to Spark programs. Identifying this ``minimum'' set is a significant result obtained through experiments that let us define the reduction module of \transmut.

\section{A comparative study on Transformation Mutation vs. Traditional Mutation Testing}\label{sec:transmut-vs-scalamu}

\transmut{} implements a transformation mutation approach to apply mutation testing in Spark programs~\cite{caise2020}. Unlike traditional mutation testing, transformation mutation considers semantic information from the Spark program, such as the dataset types and input and output types of the transformations, to modify the program's data flow and transformations. Although \transmut{} was developed for Spark programs written in Scala, the transformation mutation behind~\cite{caise2020} is independent of the programming language and the data processing system. Thus, it can be applied in programs written in other programming languages or from other data processing systems if they implement a data flow model similar to Spark, such as \textit{Apache Flink}, \textit{Apache Beam} and \textit{DryadLINQ}.

Traditional mutation testing and transformation mutation are complementary because they refer to different facets of the program and simulate different faults.
In \textit{traditional mutation testing}, modifications are made on the syntactic facet, and in consequence, it depends on the syntax of the programming language~\cite{offut:2010}. This approach mimics programming faults through syntactic deviations in the program, such as replacing one arithmetic operator with another. In contrast, transformation mutation simulates faults related to the definition of the data flow and specific transformations used in a data processing program independent of the programming language.

We experimentally validated the hypothesis that traditional mutation testing and transformation mutation are complimentary for testing  Scala programs that include data processing code relying on the Apache Spark libraries.
We experimented using a traditional mutation testing tool to compare its results with those obtained by \transmut{}.
We used the traditional mutation testing tool \textit{Scalamu}\footnote{\url{https://github.com/sugakandrey/scalamu}}  for programs written in  Scala. 
\textit{Scalamu} is based on \textit{PIT}~\cite{Coles2016}, a state of the art tool for mutation testing in Java programs.

\subsection{Methodology}

This experiment was designed to evaluate the performance of the (1) test set designed to kill the mutants generated by \transmut{} in the mutation testing process with \textit{Scalamu}; (2)  test set developed to kill the mutants generated by \textit{Scalamu} in the process with \transmut{}. 
The idea is that proof that since 
both tools (and mutation approaches) target different facets of the program, and they will lead to tests that can identify different faults.

We used the same programs applied in the experiments presented in Section~\ref{sec:experiments} and the test set designed to kill the mutants generated by \transmut{} (we refer to this test set as ``\transmut{} Tests''). Besides, 
we developed a new test set capable of killing the mutants generated by \textit{Scalamu}. We used the same methodology described in Section~\ref{sec:experiments} for designing tests. So, we developed a simple test for each program, then we performed the mutation testing process with \textit{Scalamu}, and we analyzed the living mutants to detect equivalent. Then, we developed new tests incrementally to have a test set capable of killing all the mutants generated by \textit{Scalamu} and achieve a mutation score ($ms$) of $1.0$. Therefore, we prepared a new test set capable of killing all the mutants generated by \textit{Scalamu} (we refer to this set as ``\textit{Scalamu} Tests''). Finally, we executed the two test sets with each testing tool \textit{Scalamu} and \transmut{} and compared results. 
For \transmut{}, we executed the process with the mutants reduction module enabled to remove redundant and inefficient mutants. The results of this experiment are presented below.



\subsection{Results}

Table~\ref{tab:transmut-vs-scalamu-with-transmut-2} and table~\ref{tab:transmut-vs-scalamu-with-scalamu} show the results of the experiment for comparing  \transmut{} and \textit{Scalamu}.
Table~\ref{tab:transmut-vs-scalamu-with-transmut-2} shows the results of \transmut{} with the mutants reduction module enabled. Table~\ref{tab:transmut-vs-scalamu-with-scalamu} shows the results obtained with  \textit{Scalamu}. The tables show the performance of the test set designed to kill the mutants generated by \transmut{} (``\transmut{} Tests'') and the performance of the test set designed to kill the mutants generated by \textit{Scalamu} (``\textit{Scalamu} Tests''). Besides, they show the number of tests designed for each program (\textit{Tests}), the number of generated mutants  (\textit{\#M}), the number of mutants killed (\textit{\#K}), the number of equivalent mutants (\textit{\#E}), the number of mutants removed by the reduction module (\textit{\#R}) and the mutation score (\textit{ms}). To insist on the comparison,  the mutation score values highlighted in blue indicate the cases with the same $ms$ for the two test sets.  The cases highlighted in green are those with an $ms$  of $1.0$. The cases highlighted in red are those with an $ms$  below $1.0$.

\begin{table}
\bigskip
\caption{Results obtained with \transmut{} using the reduction module for the comparative experiment of \transmut{} and \textit{Scalamu}.}
\label{tab:transmut-vs-scalamu-with-transmut-2}
\begin{adjustbox}{width=1\textwidth}
\begin{tabular}{|c|c|c|c|c|l|c|c|c|c|c|l|c|}
\hline
\multirow{2}{*}{\textbf{Program}} & \multicolumn{6}{c|}{\textbf{\transmut{} Tests}}                                         & \multicolumn{6}{c|}{\textbf{\textit{Scalamu} Tests}}                                                \\ \cline{2-13} 
                           & \textbf{Tests} & \textbf{\#M} & \textbf{\#K} & \textbf{\#E} & \textbf{\#R} & \textbf{ms}   & \textbf{Tests} & \textbf{\#M} & \textbf{\#K} & \textbf{\#E} & \textbf{\#R} & \textbf{ms}   \\ \hline
\texttt{NGramsCount}                & 5              & 23           & 22           & 1            & 9            & \cellcolor{green!30} 1.00          & 4              & 23           & 18           & 1            & 9            & \cellcolor{red!30} 0.82          \\ \hline
\texttt{ScanQuery}                  & 3              & 7            & 7            & 0            & 6            & \cellcolor{green!30} 1.00          & 1              & 7            & 1            & 0            & 6            & \cellcolor{red!30} 0.14          \\ \hline
\texttt{AggregationQuery}           & 3              & 10           & 10           & 0            & 6            & \cellcolor{green!30} 1.00          & 2              & 10           & 7            & 0            & 6            & \cellcolor{red!30} 0.70          \\ \hline
\texttt{DistinctUserVisitsPerPage}  & 2              & 11           & 7            & 4            & 6            & \cellcolor{blue!30} 1.00          & 1              & 11           & 7            & 4            & 6            & \cellcolor{blue!30} 1.00          \\ \hline
\texttt{MoviesRatingsAverage}       & 4              & 13           & 11           & 2            & 6            & \cellcolor{green!30} 1.00          & 6              & 13           & 7            & 2            & 6            & \cellcolor{red!30} 0.64          \\ \hline
\texttt{MoviesRecomendation}        & 5              & 35           & 24           & 11           & 21           & \cellcolor{blue!30} 1.00          & 9              & 35           & 24           & 11           & 21           & \cellcolor{blue!30} 1.00          \\ \hline
\texttt{JoinQuery}                  & 6              & 21           & 20           & 1            & 16           & \cellcolor{green!30} 1.00          & 2              & 21           & 7            & 1            & 16           & \cellcolor{red!30} 0.35          \\ \hline
\texttt{NasaApacheWebLogsAnalysis}  & 4              & 31           & 21           & 10           & 7            & \cellcolor{green!30} 1.00          & 3              & 31           & 15           & 10           & 7            & \cellcolor{red!30} 0.71          \\ \hline
\texttt{MovieLensExploration}       & 17             & 130          & 119          & 11           & 65           & \cellcolor{green!30} 1.00          & 10             & 130          & 92           & 11           & 65           & \cellcolor{red!30} 0.77          \\ \hline
\textbf{Total}             & \textbf{49}    & \textbf{281} & \textbf{241} & \textbf{40}  & \textbf{142} & \cellcolor{green!30} \textbf{1.00} & \textbf{38}    & \textbf{281} & \textbf{178} & \textbf{40}  & \textbf{142} & \cellcolor{red!30} \textbf{0.74} \\ \hline
\end{tabular}
\end{adjustbox}
\end{table}

\begin{table}
\bigskip
\caption{Results obtained with \textit{Scalamu} for the comparative experiment of \transmut{} and \textit{Scalamu}.}
\label{tab:transmut-vs-scalamu-with-scalamu}
\begin{adjustbox}{width=1\textwidth}
\begin{tabular}{|c|c|c|c|c|c|c|c|c|c|c|}
\hline
\multirow{2}{*}{\textbf{Program}} & \multicolumn{5}{c|}{\textbf{\transmut{} Tests}}                          & \multicolumn{5}{c|}{\textbf{\textit{Scalamu} Tests}}                                 \\ \cline{2-11} 
                           & \textbf{Tests} & \textbf{\#M} & \textbf{\#K} & \textbf{\#E} & \textbf{ms}   & \textbf{Tests} & \textbf{\#M} & \textbf{\#K} & \textbf{\#E} & \textbf{ms}   \\ \hline
\texttt{NGramsCount}                & 5              & 15           & 11           & 1            & \cellcolor{red!30} 0.79          & 4              & 15           & 14           & 1            & \cellcolor{green!30} 1.00          \\ \hline
\texttt{ScanQuery}                  & 3              & 3            & 2            & 0            & \cellcolor{red!30} 0.67          & 1              & 3            & 3            & 0            & \cellcolor{green!30} 1.00          \\ \hline
\texttt{AggregationQuery}           & 3              & 3            & 2            & 0            & \cellcolor{red!30} 0.67          & 2              & 3            & 3            & 0            & \cellcolor{green!30} 1.00          \\ \hline
\texttt{DistinctUserVisitsPerPage}  & 2              & 1            & 1            & 0            & \cellcolor{blue!30} 1.00          & 1              & 1            & 1            & 0            & \cellcolor{blue!30}1.00          \\ \hline
\texttt{MoviesRatingsAverage}       & 4              & 28           & 20           & 2            & \cellcolor{red!30} 0.77          & 6              & 28           & 26           & 2            & \cellcolor{green!30} 1.00          \\ \hline
\texttt{MoviesRecomendation}        & 5              & 44           & 34           & 1            & \cellcolor{red!30} 0.79          & 9              & 44           & 43           & 1            & \cellcolor{green!30} 1.00          \\ \hline
\texttt{JoinQuery}                  & 6              & 3            & 2            & 0            & \cellcolor{red!30} 0.67          & 2              & 3            & 3            & 0            & \cellcolor{green!30} 1.00          \\ \hline
\texttt{NasaApacheWebLogsAnalysis}  & 4              & 7            & 5            & 0            & \cellcolor{red!30} 0.71          & 3              & 7            & 7            & 0            & \cellcolor{green!30} 1.00          \\ \hline
\texttt{MovieLensExploration}       & 17             & 51           & 45           & 2            & \cellcolor{red!30} 0.92          & 10             & 51           & 49           & 2            & \cellcolor{green!30} 1.00          \\ \hline
\textbf{Total}             & \textbf{49}    & \textbf{155} & \textbf{122} & \textbf{6}   & \cellcolor{red!30} \textbf{0.82} & \textbf{38}    & \textbf{155} & \textbf{149} & \textbf{6}   & \cellcolor{green!30} \textbf{1.00} \\ \hline
\end{tabular}
\end{adjustbox}
\end{table}

\subsection{Analysis and Discussion}

As shown in the previous section, the test set's performance was different for each tool. 
The columns ``\transmut{} Tests'' in Table~\ref{tab:transmut-vs-scalamu-with-transmut-2}, and ``\textit{Scalamu} Tests'' in Table~\ref{tab:transmut-vs-scalamu-with-scalamu} show that both tools achieved a mutation score $1.0$. This score value was expected, given that we developed a specific test set  for each tool. In contrast, note that the mutation score (ms) of the test set designed for \textit{Scalamu} achieved a lower score values by \transmut{} and vice-versa.
The test set designed to kill the mutants generated by \textit{Scalamu} achieved an average ms of $0.74$, killing \transmut{} mutants generated with the reduction module enabled. The test set designed to kill the mutants generated by \transmut{} achieved an average ms of $0.82$  killing \textit{Scalamu} generated mutants. 

These results show that tests designed to kill \transmut{} mutants did not kill all of \textit{Scalamu} mutants, and vice-versa. 
These results confirm the hypothesis that both mutation approaches are complementary, and they are needed to thoroughly test a Scala program weaving code for processing data using Spark libraries operations.

\section{Threats to validity and Limitations}\label{sec:threats}

\noindent
{\em Threats to validity.}
The experimental validation presented in this work
was the first attempt to evaluate our approach and tool. However, the experiments have some limitations and threats to their validity. The first limitation is related to the testing battery.   Programs implement code that focuses on the pure Big Data processing tasks within applications because \transmut{} tests this type of code and not the code that wraps these tasks. 
This choice implies that the programs of the testing battery are 
simple in terms of the number of lines, program logic, and scope, giving the impression of not being adapted for testing complex, industrial applications. Nevertheless, the complexity of the programs in the testing could increase concatenating several Big Data processing tasks. However,  this strategy does not seem realistic because, usually, applications do not combine several analytics targets at a time. Besides, the phases of the analytics process like data preparation, cleaning, training and applying models~\cite{zoller2021benchmark} are treated as separated programs, just as we did with our testing battery.

A second limitation for our study is related to the tests designed to kill mutants in experiments. 
Our goal in the experiments was to design tests that would kill all generated non-equivalent mutants. 
Thus, experiments provide a quantitative profile of the approach and its automation. 
The experiments are not defined to assess the approach and tool concerning other criteria (such as graph-based or input partitioning coverage~\cite{offut:2010}), and they do not provide insight to compare the results against other testing techniques.
Also, the fact that the authors themselves develop the tests in the experiments may usually introduce a bias in the evaluation. 
We mitigated this threat by building the test set from simple tests as a starting point and evolving minimally until reaching a mutation score of $1.0$. 
We managed to avoid having unnecessary tests, and more importantly, unnecessarily complex tests so that the developed tests all contributed to the testing process and, at the same time, were not more powerful than required by the process. 



\noindent
{\em Rationale of limitations and outreach.}
\transmut{} automates the mutation testing process for testing Apache Spark programs, 
relying on mutation operators that tackle faults (see Section~\ref{sec:mutation_operators}) that can come up due to (1) wrong data flow definition; (2) inappropriate use of transformations which are Spark built-in operations for processing data that must be chosen by programmers according to their semantics and use specifications; and (3) variables used to guide the data sharing across distributed parallel processes. \transmut{} also relies on reduction rules that apply a selective mutation approach~\cite{Rothermel1993} to remove redundant and inefficient mutants.
\transmut{} uses these rules in a reduction module that prunes the generated mutants and produces a ``minimum'' set to be used in the testing process.
The assessment scores obtained experimentally (mutation score, killed ratio and process/execution time) showed promising results of \transmut{} in the process of testing Spark data processing programs.

The current version of \transmut{} relies on a simple strategy for managing test cases and mutants. Regarding tests, \transmut{} operates at the test class level, where a class can contain the implementation of one or more test cases. Therefore, the tool allows specifying which test classes should be executed or not in the testing process. However, it cannot deal with a finer granularity selecting test cases within classes. 

Regarding mutants, it is possible to define which operators will be applied or not in the process with \transmut{}, but it does not allow to select and execute individual mutants.
The reason is that \transmut{} executes all the testing steps sequentially within a unique process (see Figure~\ref{fig:transmut-process}). The process is not interactive to let calibrate steps according to partial results—for example,  select mutants to be executed between the generation and execution steps.
Still,
\transmut{}  can execute a subset of mutants that survived a previous testing process.  It also lets to tag equivalent mutants to avoid unnecessary executions.   \transmut{} allows tests to be added incrementally to the process, such that new tests can be developed to kill mutants that previous tests could not kill. These functionalities reduce the process execution time because only the necessary mutants are executed.

The program code \transmut{} can only test Spark programs built using the patterns it supports. Otherwise, programs have to be refactored before they are tested with \transmut{}.  We decided to support specific patterns for the first version of \transmut{} to facilitate 
(1)  \textit{controllability} because the program can  run independently of the context; 
(2) \textit{observability} of the program  elements (datasets and transformation) and behavior in the tests. These characteristics are fundamental to enable the automation of tests~\cite{offut:2010}.

\section{Conclusions and Future Work}\label{sec:conclusion}

This paper introduced \transmut, a transformation mutation tool for Spark large-scale data processing programs.  
\transmut{} implements operators for mutating the data flow and the transformations composing a Spark program~\cite{caise2020}. 
Through the description of the tool and experiments, we showed that  \transmut{} deals with \textit{test case handling} with the possibility for executing, including, and excluding test cases.  
\transmut{} also deals with \textit{mutant handling}, by generating, executing, and analyzing mutants. Finally,  \transmut{}  performs an adequacy analysis, calculating the mutation score and generating reports.  
These are the main requirements for a mutation testing tool~\cite{delamaro1996proteum}.
In this way, \transmut{} automates the principal and most laborious steps of the mutation testing process; thereby, it fully executes the process in Spark programs. 
\transmut{} addresses mutation testing for the data processing aspects of programs. Therefore, we decided to show that it is complementary to classical mutation testing tools. Indeed, experiments show that \transmut{} and \textit{Scalamu} combined can lead to validation of both Scala code and weaved Spark data processing code. 
The current version of \transmut\  is available on Github, and it can be used to test Spark data processing programs.

For the time being, the experiments run on  \transmut{} have addressed representative data processing code, validating the mutation operators proposed in a previous work~\cite{caise2020}. 
Experimental results confirmed the tool's utility, even if we can further use other testing batteries with programs implementing a more significant number of processing tasks and compare results with other testing approaches. We will also tackle testing classic programs weaving data processing code using
classical testing techniques (such as input space partitioning and logical coverage~\cite{offut:2010}) and with other work in the field such as~\cite{riesco2015}. 

Finally, the mutation operators~\cite{caise2020} were formalized with the model for data flow programs presented in~\cite{sbmf2020},  based on characteristics of different systems, including Apache Flink~\cite{katsifodimos2015apache}, Apache Beam~\cite{beam2016}, and DryadLINQ~\cite{yu2008}. 
Thus, we also plan to extend \transmut{} to apply mutation testing to programs in other systems besides Spark.

\ack 
This study was partially funded by the Coordena\c c\~ ao de Aperfei\c coamento de Pessoal de N\'\i vel Superior - Brasil (CAPES) - Finance Code 001.

\bibliographystyle{wileyj}
\bibliography{references.bib}

\end{document}